\documentclass{aa}
\usepackage{graphicx}
\usepackage{rotating,times,pictex,graphicx,latexsym}
\usepackage{color}
\begin{document}
\title{Star formation in globules in IC\,1396}

\author{Dirk Froebrich$^1$\thanks{Visiting Astronomer at the German-Spanish
Astronomical Centre, Calar Alto, operated by the Max-Planck-Institut f\"ur
Astronomie, Heidelberg, jointly with the Spanish National Commission for
Astronomy.}, Alexander Scholz$^2$, Jochen Eisl\"offel$^2$ \and Gareth\,C.
Murphy$^1$} 

\offprints{df@cp.dias.ie}

\institute{$^1$ Dublin Institute for Advanced Studies, 5 Merrion Square, Dublin
2, Ireland \\ $^2$ Th\"uringer Landessternwarte Tautenburg, Sternwarte 5,
D-07778 Tautenburg, Germany}

\authorrunning{D.~Froebrich et al.}

\date{Received sooner ; accepted later}

\abstract{We present a large-scale study of the \object{IC\,1396} region using
new deep NIR and optical images, complemented by 2MASS data. For ten globules
in IC\,1396 we determine (H-K, J-H) colour-colour diagrams and identify the
young stellar population. Five of these globules contain a rich population of
reddened objects, most of them probably young stellar objects. Two new HH 
objects (\object{HH\,865} and \object{HH\,864}) could be identified by means of
[SII] emission, one of  them a parsec-scale flow. Using star counts based on
2MASS data we create an extinction map of the whole region. This map is used to
identify 25 globules and to estimate their mass. The globule masses show a
significant increase with the distance from the exciting O6.5V star
\object{HD\,206267}. We  explain this correlation by the enhanced radiation
pressure close to this star,  leading to evaporation  of the nearby clouds and
hence smaller globule masses. We  see evidence that the radiation from
HD\,206267 has a major impact on the star formation activity in these globules.

}
\maketitle
%

\section{Introduction}

Star formation takes place not only in giant molecular clouds but also in small
isolated globules. Works e.g. by Sugitani et al. (\cite{1991ApJS...77...59S}),
Schwartz et al. (\cite{1991ApJ...370..263S}) and others showed that such
places are associated with young stellar objects. The identification of young
stellar clusters or outflow activity in such globules hints an ongoing star
formation process. Star formation in globules might be induced by the
propagation of an ionisation shock front, the so-called radiation driven
implosion mechanism (Reipurth \cite{1983A&A...117..183R}). However, it is not
fully clear what the main properties are that influence star formation within
these globules (e.g. density, mass, or size of the globule, strength of the
ionisation shock front). Further, it would be interesting to see if and how
these properties influence the number and clustering of the forming stars. The
investigation of a larger, homogeneous, and as far as possible unbiased sample
of globules is an ideal way to obtain a deeper understanding of these issues. 

IC\,1396 is one of the youngest and most active H\,II regions in the
\object{Cep\,OB\,2} group of loosely clustered OB stars (Schwartz et al.
\cite{1991ApJ...370..263S}). The nebula is excited by the O6.5V star HD\,206267
(Walborn \& Panek \cite{1984ApJ...286..718W}) and contains 15 small clouds and
globules associated with red IRAS sources (Schwartz et al.
\cite{1991ApJ...370..263S}) at a distance of about 750\,pc (Matthews
\cite{1979A&A....75..345M}). Large-scale observations of this region in the
rotational CO lines were done by Patel et al. (\cite{1995ApJ...447..721P}) and
Weikard et al. (\cite{1996A&A...309..581W}). The numerous sharp-rimmed clouds
and the relative proximity make this region an ideal place to study star
formation in a large, homogeneous sample of small globules. 

Some of the globules in IC\,1396 were already investigated in detail
(\object{IC\,1396\,N}: Beltran et al. \cite{2002ApJ...573..246B}, Nisini et al.
\cite{2001A&A...376..553N}, Codella et al. \cite{2000ApJ...542..464C};
\object{IC\,1396\,W}: Froebrich \& Scholz \cite{2003A&A...407..207F};
\object{IC\,1396\,A} or the Elephant Trunk Nebula: Nakano et al.
\cite{1989PASJ...41.1073N}, Hessman et al. \cite{1995A&A...299..464H}, Reach et
al. \cite{2004ApJS..154..385R}) and/or are known to harbour outflow sources
(\object{IRAS\,21388+5622}, Duvert et al. \cite{1990A&A...233..190D}, Sugitani
et al. \cite{1997ApJ...486L.141S}, De Vries et al. \cite{2002ApJ...577..798D},
Ogura et al. \cite{2002AJ....123.2597O}; \object{IRAS\,22051+5848}, Reipurth \&
Bally \cite{2001ARA&A..39..403R}). Relatively little is known about most of the
other globules, which are therefore targets for this new  study. As shown e.g.
in Froebrich \& Scholz (\cite{2003A&A...407..207F}), deep NIR imaging in JHK
and the construction of (H-K, J-H) colour-colour diagrams can reveal embedded
young objects or clusters in such globules. These data can be used in
conjunction with star counts (see e.g. Kiss et al. \cite{2000A&A...363..755K})
to estimate the extinction and hence determine the mass of the globules.
Simultaneous observations of the region in narrow-band filters centred either
on the 1-0\,S(1) line of molecular hydrogen at 2.122\,$\mu$m or the [SII] lines
at 671.6 and 673.1\,nm will uncover outflow activity from young stars.

This paper is structured as follows: In Sect.\,\ref{observations} we describe
our data obtained in the optical and NIR and specify the data reduction
process. Our photometry in the NIR images with emphasis on (H-K, J-H)
colour-colour diagrams is shown in Sect.\,\ref{photometry}. The creation of NIR
extinction maps to estimate the mass of the globules is described in
Sect.\,\ref{extmaps}, followed by a discussion of the determined globule
properties in Sect.\,\ref{activity} and the characterisation of the newly
discovered outflows in Sect.\,\ref{outflows}. Finally, the results are
discussed and summarised in Sect.\,\ref{discuss}. Some more detailed technical
descriptions of the data analysis procedures can be found in
Appendices\,\ref{reliability}\,-\,\ref{exmade}.

\section{Observations and data reduction} 

\label{observations}

\subsection{Near infrared data}
\label{nirobs}

The near infrared (NIR) observations were made towards 10 out of the 15 IRAS
sources listed in Schwartz et al. (\cite{1991ApJ...370..263S}). Two of them
have been studied before (IC\,1396\,N, Beltran et al.
\cite{2002ApJ...573..246B}, Nisini et al. \cite{2001A&A...376..553N}, Codella
et al. \cite{2000ApJ...542..464C}; IC\,1396\,W Froebrich \& Scholz
\cite{2003A&A...407..207F}), and another three globules could not be observed
due to bad weather conditions during our run. The observed globules are listed
on the top of Table\,\ref{globules}. One main purpose of the NIR data was to
construct colour-colour diagrams for each globule in which we want to detect
reddened sources. This can be done by comparing the measured colours of the
targets with that of the main sequence. To get an estimate for the colours of
the main sequence, we observed  main sequence stars with known spectral type in
the same way as our IC\,1396 targets. These standard stars were selected from
the SIMBAD database and deliver an unreddened main sequence for our
colour-colour diagrams. We preferred to define this main sequence by
observations rather than with theoretical evolutionary tracks, because this
avoids mismatches from inconsistent photometric systems, which can be
substantial, as we have shown in Froebrich \& Scholz
(\cite{2003A&A...407..207F}). We selected the  standard stars so that they are
generally less than 10\,degrees away from the IC\,1396 region. Thus, they can
be observed roughly at the same airmass as IC\,1396, which excludes
systematic offsets because of differential extinction.

Our near infrared data were obtained on six nights from the 18th to 23rd of
July in 2003 with the 2.2-m telescope on Calar Alto, Spain. We observed with
the MAGIC camera (Herbst et al. \cite{1993SPIE.1946..605H}) in its wide-field
mode (6\arcmin$\!\!$.92\,$\times$\,6\arcmin$\!\!$.92 FoV). Using a 3\,x\,3
dither pattern around the central coordinates with a shift of half a detector
size we obtained 13\arcmin$\!\!$.5\,$\times$\,13\arcmin$\!\!$.5 sized mosaics
of the IC\,1396 globules. This field size is comparable with the typical
diameter of small globules in this region (see Table\,\ref{globules}).

For the standard star observations, we chose a smaller shift between the images
to reduce overhead times. All fields were observed  in J, H, and K$'$, the
IC\,1396 globules additionally in a narrow-band filter centred on the 1-0\,S(1)
line of molecular hydrogen at 2.122\,$\mu$m (hereafter called the
H$_2$-filter). The K$'$ filter is very similar to the 2MASS Ks band and will be
called K in the following. The total per-pixel integration time in the IC\,1396
mosaics was 324\,s in each broad-band filter. The integration times  for the
H$_2$-filter were at least 2160\,s per pixel, in most cases we reached  3240\,s
per pixel. For the standard stars, appropriate integration times were  chosen
to avoid detector saturation. 

\begin{figure*}[t]
\beginpicture
\setcoordinatesystem units <12.58mm,22.563mm> point at 100 0
\setplotarea x from 319.5 to 333 , y from 56 to 59.5
\put {\includegraphics[angle=0,width=170mm]{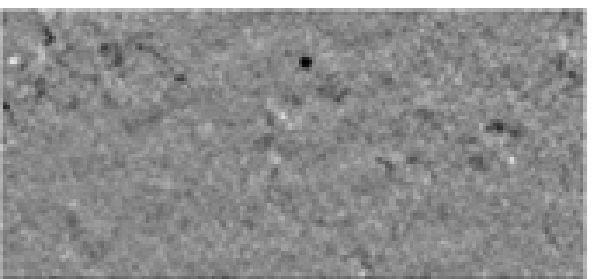}} at 326.25 57.75
{\color{white}
\axis left label {}
ticks in long unlabeled from 56 to 59.5 by 0.5
      short unlabeled from 56 to 59.5 by 0.1 /
\axis right label {}
ticks in long unlabeled from 56 to 59.5 by 0.5
      short unlabeled from 56 to 59.5 by 0.1 /
\axis bottom label {}
ticks in long unlabeled from 321.0 to 331.5 by 1.5
      short unlabeled from 320 to 333 by 0.5 /
\axis top label {}
ticks in long unlabeled from 321.0 to 331.5 by 1.5
      short unlabeled from 320 to 333 by 0.5 /
\setdashes <1mm>
\plot 
326.0 58.75 331.0 58.75
331.0 56.25 330.0 56.25
330.0 55.75 327.0 55.75
327.0 56.25 326.0 56.25
326.0 56.44 324.7 56.44
324.7 57.58 322.5 57.58
322.5 59.58 325.1 59.58
325.1 58.44 326.0 58.44
326.0 58.75 /
\plot
320.1 58.80 321.4 58.80
321.4 60.13 320.1 60.13
320.1 58.80 /
\setsolid
\plot
320.5875 59.1586 320.9902 59.1586
320.9902 58.9336 320.5875 58.9336
320.5875 59.1586 /
\put {11} at 320.45 59.0
\plot
323.4138 58.4822 323.4138 58.7072
323.8162 58.7072 323.8162 58.4822
323.4138 58.4822 /
\put {10} at 323.25 58.6
\plot
326.2026 58.1667 326.2026 58.3917
326.605 58.3917 326.605 58.1667
326.2026 58.1667 /
\put {8} at 326.1 58.3
\plot
325.7705 57.3272 325.7705 57.5522
326.1729 57.5522 326.1729 57.3272
325.7705 57.3272 /
\put {9} at 325.7 57.4
\plot
327.178 56.4869 327.178 56.7119
327.5804 56.7119 327.5804 56.4869
327.178 56.4869 /
\put {7} at 327.1 56.6
\plot
328.1 57.3792 328.1 57.6042
328.5024 57.6042 328.5024 57.3792
328.1 57.3792 /
\put {5} at 328.5 57.27
\plot
328.2471 57.3469 328.2471 57.5719
328.6495 57.5719 328.6495 57.3469
328.2471 57.3469 /
\put {4} at 328.15 57.3
\plot
328.7959 57.38 328.7959 57.605
329.1983 57.605 329.1983 57.38
328.7959 57.38 /
\put {3} at 329.0 57.3
\plot
329.1155 57.7181 329.1155 57.9431
329.5179 57.9431 329.5179 57.7181
329.1155 57.7181 /
\put {2} at 329.0 57.8
\plot
328.0559 58.4975 328.0559 58.7225
328.4583 58.7225 328.4583 58.4975
328.0559 58.4975 /
\put {6} at 327.95 58.6
\plot
330.7713 57.8256 330.7713 58.0506
331.1737 58.0506 331.1737 57.8256
330.7713 57.8256 /
\put {1} at 330.7 57.9

\put {\circle{15}} at 331.55 57.88
\put {12} at 331.65 57.88
\put {\circle{15}} at 331.35 58.61
\put {13} at 331.45 58.61
\put {\circle{15}} at 330.70 57.52
\put {14} at 330.80 57.52
\put {\circle{15}} at 329.35 58.15
\put {16} at 329.45 58.15
\put {\circle{15}} at 328.10 56.22
\put {17} at 328.20 56.22
\put {\circle{15}} at 327.50 58.33
\put {19} at 327.60 58.33
\put {\circle{15}} at 325.85 57.77
\put {21} at 325.95 57.77
\put {\circle{15}} at 325.45 56.72
\put {22} at 325.55 56.72
\put {\circle{22}} at 322.05 58.80
\put {23} at 322.55 58.90
\put {\circle{15}} at 320.60 58.38
\put {24} at 320.70 58.38
\put {\circle{15}} at 320.60 58.52
\put {25} at 320.70 58.52

\put {+} at 326.309 57.9064
\put {+} at 326.479 58.0297
\put {+} at 326.039 58.1067
\put {+} at 326.032 57.8683
\put {+} at 325.93 58.0619
\put {+} at 326.543 57.89
\put {+} at 325.872 58.0189
\put {+} at 326.203 57.8006
\put {+} at 326.437 57.8086
\put {+} at 325.988 57.8336
\put {+} at 326.34 58.2219
\put {+} at 325.871 57.8533
\put {+} at 325.812 58.1033
\put {+} at 325.887 57.8103
\put {+} at 326.332 58.2708
\put {+} at 326.404 58.2792
\put {+} at 326.533 58.2522
\put {+} at 325.778 57.8106
\put {+} at 326.852 58.0189
\put {+} at 326.365 58.3183
\put {+} at 326.854 58.0678
\put {+} at 326.185 57.6531
\put {+} at 325.922 58.3147
\put {+} at 325.775 57.7311
\put {+} at 326.856 58.1903
\put {+} at 326.433 57.6275
\put {+} at 325.582 58.1633
\put {+} at 325.633 57.7758
\put {+} at 325.745 57.6906
\put {+} at 326.96 58.2081
\put {+} at 326.896 57.7344
\put {+} at 327.104 58.0589
\put {+} at 325.385 58.02
\put {+} at 326.994 57.7608
\put {+} at 327.062 58.2181
\put {+} at 326.409 58.4742
\put {+} at 327.159 57.9769
\put {+} at 326.684 58.4261
\put {+} at 325.278 58.1486
\put {+} at 325.584 57.5953
\put {+} at 325.413 57.6914
\put {+} at 325.952 58.5214
\put {+} at 326.943 58.405
\put {+} at 326.78 58.4781
\put {+} at 325.972 57.4397
\put {+} at 326.925 58.4725
\put {+} at 325.819 57.4539
\put {+} at 325.222 58.2411
\put {+} at 327.279 58.2633
\put {+} at 326.61 57.4244
\put {+} at 326.511 58.5961
\put {+} at 327.424 58.0397
\put {+} at 327.323 58.2694
\put {+} at 325.995 57.3864
\put {+} at 327.044 58.4694
\put {+} at 325.858 58.6017
\put {+} at 327.008 58.4986
\put {+} at 327.311 58.3314
\put {+} at 325.928 57.3717
\put {+} at 325.293 58.4336
\put {+} at 326.41 58.6622
\put {+} at 327.188 58.4517
\put {+} at 327.327 58.3667
\put {+} at 326.863 58.5944
\put {+} at 325.344 57.5233
\put {+} at 325.371 58.5064
\put {+} at 325.392 58.5219
\put {+} at 327.244 58.4528
\put {+} at 325.534 58.5917
\put {+} at 326.427 57.295
\put {+} at 326.891 57.3725
\put {+} at 325.878 57.3114
\put {+} at 327.563 58.2458
\put {+} at 327.472 58.3567
\put {+} at 327.284 58.5022
\put {+} at 326.858 57.3347
\put {+} at 325.527 57.3658
\put {+} at 327.044 58.6181
\put {+} at 324.825 57.935
\put {+} at 325.976 57.2503
\put {+} at 325.773 57.2717
\put {+} at 327.556 57.6594
\put {+} at 325.349 57.3897
\put {+} at 325.199 57.4544
\put {+} at 324.818 58.2017
\put {+} at 326.043 58.7867
\put {+} at 327.724 57.8586
\put {+} at 326.622 58.7794
\put {+} at 326.183 57.1828
\put {+} at 325.847 57.2069
\put {+} at 327.617 58.4036
\put {+} at 327.722 57.7308
\put {+} at 324.714 57.8403
\put {+} at 324.907 58.4428
\put {+} at 324.722 57.7786
\put {+} at 325.153 57.3914
\put {+} at 326.281 58.8544
\put {+} at 325.939 57.1497
\put {+} at 326.772 58.8275
\put {+} at 325.367 58.7417
\put {+} at 325.971 57.1222
\put {+} at 327.636 57.4786
\put {+} at 328.002 58.2233
\put {+} at 324.482 57.8214
\put {+} at 324.713 57.5078
\put {+} at 327.556 57.3394
\put {+} at 325.326 57.1703
\put {+} at 326.095 57.0344
\put {+} at 327.69 57.4061
\put {+} at 327.824 57.505
\put {+} at 327.785 57.4606
\put {+} at 324.462 57.7381
\put {+} at 328.052 57.7375
\put {+} at 324.576 57.555
\put {+} at 326.801 58.9553
\put {+} at 327.711 58.6458
\put {+} at 327.926 57.5219
\put {+} at 325.826 57.0044
\put {+} at 324.596 57.4847
\put {+} at 325.283 58.8919
\put {+} at 325.773 57.0047
\put {+} at 327.461 57.1944
\put {+} at 325.878 59.0214
\put {+} at 327.892 57.4044
\put {+} at 327.521 58.8261
\put {+} at 327.161 57.0567
\put {+} at 328.087 57.5864
\put {+} at 326.024 59.0583
\put {+} at 327.963 57.4467
\put {+} at 328.078 57.5583
\put {+} at 327.689 58.7683
\put {+} at 325.059 57.1269
\put {+} at 326.238 56.8997
\put {+} at 328.293 57.8106
\put {+} at 327.253 58.9772
\put {+} at 325.128 57.0636
\put {+} at 326.056 56.8728
\put {+} at 326.636 56.8814
\put {+} at 325.24 56.9989
\put {+} at 324.211 57.65
\put {+} at 327.857 57.2486
\put {+} at 324.3 57.5281
\put {+} at 328.074 58.6228
\put {+} at 328.212 58.4947
\put {+} at 328.172 58.5403
\put {+} at 328.37 58.2628
\put {+} at 324.518 57.3031
\put {+} at 324.411 58.6436
\put {+} at 326.983 56.9069
\put {+} at 324.073 58.1694
\put {+} at 326.238 56.8325
\put {+} at 325.582 56.8836
\put {+} at 325.28 56.9339
\put {+} at 324.463 57.2908
\put {+} at 327.859 57.1647
\put {+} at 328.301 57.4917
\put {+} at 325.879 56.8011
\put {+} at 328.257 58.61
\put {+} at 324.947 56.99
\put {+} at 328.489 57.6889
\put {+} at 324.039 57.6356
\put {+} at 328.576 57.8119
\put {+} at 326.43 59.2519
\put {+} at 325.09 56.9086
\put {+} at 323.903 57.8144
\put {+} at 325.618 56.7769
\put {+} at 323.923 57.7142
\put {+} at 328.501 58.4617
\put {+} at 323.954 58.3922
\put {+} at 326.023 59.2722
\put {+} at 328.323 58.6703
\put {+} at 328.477 58.5275
\put {+} at 323.903 57.695
\put {+} at 328.448 57.4594
\put {+} at 325.572 56.7603
\put {+} at 323.827 57.8725
\put {+} at 326.251 56.7061
\put {+} at 326.81 59.2703
\put {+} at 327.102 56.7761
\put {+} at 328.706 58.2381
\put {+} at 326.702 59.2997
\put {+} at 328.61 58.4536
\put {+} at 327.341 56.8106
\put {+} at 324.778 56.9339
\put {+} at 328.75 57.9125
\put {+} at 328.552 58.5531
\put {+} at 323.928 58.5561
\put {+} at 328.671 57.6097
\put {+} at 323.749 57.7506
\put {+} at 327.899 56.9767
\put {+} at 326.546 56.6489
\put {+} at 324.033 58.7064
\put {+} at 328.821 58.0772
\put {+} at 327.052 59.2986
\put {+} at 328.48 57.3347
\put {+} at 324.243 57.1536
\put {+} at 325.396 56.7022
\put {+} at 323.637 57.88
\put {+} at 324.024 57.2464
\put {+} at 327.546 56.7725
\put {+} at 328.66 57.4186
\put {+} at 323.575 57.975
\put {+} at 328.778 58.495
\put {+} at 324.417 56.9675
\put {+} at 324.832 59.2308
\put {+} at 323.559 57.8522
\put {+} at 328.949 58.1753
\put {+} at 325.452 56.6272
\put {+} at 326.22 56.56
\put {+} at 323.795 57.4006
\put {+} at 324.071 58.8858
\put {+} at 328.859 58.4592
\put {+} at 328.773 58.6042
\put {+} at 326.796 56.5675
\put {+} at 323.56 57.6558
\put {+} at 323.784 57.33
\put {+} at 328.774 57.3906
\put {+} at 324.372 59.1153
\put {+} at 323.452 58.1122
\put {+} at 329.033 58.2344
\put {+} at 328.339 57.0183
\put {+} at 326.832 59.4656
\put {+} at 328.257 56.9619
\put {+} at 323.438 57.845
\put {+} at 323.615 58.5947
\put {+} at 329.111 58.0989
\put {+} at 327.108 56.5553
\put {+} at 329.115 58.0578
\put {+} at 323.376 58.0264
\put {+} at 327.379 56.5994
\put {+} at 323.543 57.4958
\put {+} at 324.495 56.7919
\put {+} at 326.653 56.4808
\put {+} at 329.146 58.1406
\put {+} at 324.694 56.7089
\put {+} at 327.32 59.4417
\put {+} at 328.158 56.8414
\put {+} at 328.665 57.1492
\put {+} at 328.997 57.4925
\put {+} at 323.91 57.0869
\put {+} at 324.59 59.2936
\put {+} at 323.99 57.0258
\put {+} at 328.976 57.4344
\put {+} at 324.412 56.7889
\put {+} at 324.487 56.7572
\put {+} at 329.175 57.8039
\put {+} at 326.967 56.4817
\put {+} at 329.207 58.0786
\put {+} at 329.066 57.5294
\put {+} at 326.691 56.4442
\put {+} at 329.233 57.9689
\put {+} at 324.059 59.0992
\put {+} at 327.05 56.47
\put {+} at 325.105 59.4856
\put {+} at 329.269 58.2642
\put {+} at 329.261 58.3119
\put {+} at 323.227 58.2981
\put {+} at 323.937 56.9564
\put {+} at 323.765 58.9853
\put {+} at 325.684 56.3931
\put {+} at 329.317 57.8306
\put {+} at 323.901 59.1031
\put {+} at 329.199 57.4928
\put {+} at 324.753 59.4603
\put {+} at 324.345 56.7028
\put {+} at 327.357 56.4453
\put {+} at 323.682 57.0328
\put {+} at 323.42 58.8064
\put {+} at 329.378 57.7044
\put {+} at 324.262 56.6642
\put {+} at 323.064 58.3517
\put {+} at 324.076 59.2936
\put {+} at 329.49 57.9583
\put {+} at 323.879 56.8475
\put {+} at 328.769 56.9397
\put {+} at 328.879 57.0169
\put {+} at 323.519 58.9631
\put {+} at 329.468 57.7842
\put {+} at 323.098 57.5944
\put {+} at 328.392 59.3294
\put {+} at 324.951 56.4019
\put {+} at 328.307 56.6492
\put {+} at 323.047 58.4806
\put {+} at 323.939 56.7594
\put {+} at 329.573 58.0214
\put {+} at 323.328 58.8853
\put {+} at 324.566 56.4753
\put {+} at 326.106 56.2242
\put {+} at 323.393 58.965
\put {+} at 328.531 56.7139
\put {+} at 329.598 57.8589
\put {+} at 324.218 56.5906
\put {+} at 329.401 57.4022
\put {+} at 327.96 56.4653
\put {+} at 324.311 56.5447
\put {+} at 325.004 56.335
\put {+} at 329.639 57.8711
\put {+} at 324.636 56.4172
\put {+} at 329.308 57.2181
\put {+} at 323.061 57.3561
\put {+} at 324.2 56.5492
\put {+} at 328.775 56.7725
\put {+} at 324.395 56.4592
\put {+} at 329.108 56.9589
\put {+} at 322.87 57.5369
\put {+} at 327.949 56.3814
\put {+} at 329.605 57.4744
\put {+} at 322.765 58.3303
\put {+} at 327.635 56.285
\put {+} at 325.671 56.1467
\put {+} at 323.314 56.9725
\put {+} at 323.414 59.1681
\put {+} at 325.06 56.2289
\put {+} at 322.757 58.4714
\put {+} at 323.69 56.69
\put {+} at 329.355 57.0794
\put {+} at 329.451 57.1681
\put {+} at 323.861 56.5886
\put {+} at 329.53 57.24
\put {+} at 326.502 56.09
\put {+} at 322.771 58.5889
\put {+} at 323.747 56.6331
\put {+} at 327.038 56.1258
\put {+} at 328.088 56.3556
\put {+} at 323.231 59.1053
\put {+} at 323.777 56.6108
\put {+} at 329.878 58.2383
\put {+} at 324.087 56.4628
\put {+} at 328.613 56.5431
\put {+} at 322.595 58.2017
\put {+} at 329.906 58.2172
\put {+} at 325.224 56.1381
\put {+} at 328.283 56.3928
\put {+} at 329.747 58.6542
\put {+} at 328.952 56.7053
\put {+} at 322.703 57.5247
\put {+} at 322.566 57.9686
\put {+} at 324.781 56.2131
\put {+} at 322.766 57.3844
\put {+} at 323.052 59.0353
\put {+} at 329.932 58.2939
\put {+} at 323.964 56.4719
\put {+} at 329.963 58.1469
\put {+} at 329.904 58.4808
\put {+} at 326.68 56.0281
\put {+} at 323.087 59.1144
\put {+} at 329.833 57.4272
\put {+} at 326.082 56.0028
\put {+} at 322.669 57.3961
\put {+} at 329.758 57.2881
\put {+} at 329.154 56.7431
\put {+} at 322.621 58.6319
\put {+} at 323.2 59.2303
\put {+} at 328.332 56.3281
\put {+} at 322.848 58.9525
\put {+} at 329.69 57.1797
\put {+} at 322.903 59.0208
\put {+} at 329.817 58.7792
\put {+} at 323.251 56.7714
\put {+} at 328.48 56.3669
\put {+} at 329.715 58.9208
\put {+} at 328.945 56.5733
\put {+} at 327.752 56.1267
\put {+} at 330.121 57.99
\put {+} at 325.035 56.0478
\put {+} at 323.372 56.635
\put {+} at 330.134 57.8256
\put {+} at 322.341 58.1183
\put {+} at 328.866 56.4792
\put {+} at 327.509 56.0383
\put {+} at 323.252 59.3772
\put {+} at 329.665 57.0094
\put {+} at 322.532 57.3633
\put {+} at 322.585 57.2717
\put {+} at 329.961 57.3342
\put {+} at 322.797 59.0731
\put {+} at 328.115 56.1631
\put {+} at 324.219 56.2117
\put {+} at 323.145 59.3647
\put {+} at 322.263 58.1731
\put {+} at 322.789 59.1183
\put {+} at 329.146 56.5592
\put {+} at 324.78 56.0058
\put {+} at 322.993 56.7469
\put {+} at 323.364 56.5036
\put {+} at 322.22 58.4272
\put {+} at 324.389 56.0858
\put {+} at 329.868 57.0369
\put {+} at 322.363 58.7894
\put {+} at 330.193 58.7031
\put {+} at 322.217 57.5894
\put {+} at 322.5 57.1297
\put {+} at 330.18 57.3533
\put {+} at 323.702 56.2881
\put {+} at 328.745 56.2575
\put {+} at 323.641 56.2836
\put {+} at 329.862 59.1875
\put {+} at 323.078 56.5736
\put {+} at 322.877 56.7056
\put {+} at 328.935 56.3044
\put {+} at 323.332 56.4064
\put {+} at 323.494 56.3239
\put {+} at 322.876 59.4125
\put {+} at 322.013 58.2403
\put {+} at 322.475 57.0083
\put {+} at 323.265 56.4192
\put {+} at 321.979 58.0014
\put {+} at 322.011 58.355
\put {+} at 330.46 58.4708
\put {+} at 323.473 56.3
\put {+} at 330.341 58.7547
\put {+} at 322.012 58.4528
\put {+} at 322.602 56.8092
\put {+} at 330.3 58.8908
\put {+} at 329.218 56.3536
\put {+} at 329.421 56.4431
\put {+} at 323.524 56.2164
\put {+} at 329.132 56.2803
\put {+} at 323.87 56.0656
\put {+} at 321.87 57.8047
\put {+} at 322.613 59.3775
\put {+} at 329.007 56.1978
\put {+} at 321.805 58.1569
\put {+} at 330.701 58.2269
\put {+} at 330.638 57.6519
\put {+} at 330.689 57.7581
\put {+} at 330.72 58.3486
\put {+} at 330.563 57.3856
\put {+} at 323.313 56.2167
\put {+} at 330.116 59.3106
\put {+} at 321.717 57.7792
\put {+} at 321.802 57.5133
\put {+} at 330.225 59.2642
\put {+} at 321.665 58.0678
\put {+} at 323.249 56.1881
\put {+} at 322.094 59.1194
\put {+} at 322.448 59.4478
\put {+} at 330.627 57.2614
\put {+} at 330.892 58.1178
\put {+} at 330.671 58.8278
\put {+} at 330.021 56.5989
\put {+} at 322.331 56.7103
\put {+} at 322.05 56.97
\put {+} at 321.715 57.4489
\put {+} at 321.647 57.5578
\put {+} at 330.773 57.3731
\put {+} at 321.527 58.1808
\put {+} at 330.974 57.9378
\put {+} at 330.921 58.5142
\put {+} at 322.909 56.2628
\put {+} at 330.009 56.48
\put {+} at 321.455 58.1333
\put {+} at 330.559 56.9428
\put {+} at 331.072 58.1789
\put {+} at 329.379 56.0894
\put {+} at 322.373 56.5322
\put {+} at 322.313 56.5736
\put {+} at 330.971 58.6994
\put {+} at 321.668 57.1986
\put {+} at 330.681 59.1767
\put {+} at 330.74 56.9983
\put {+} at 331.2 58.165
\put {+} at 321.278 58.065
\put {+} at 321.333 58.5069
\put {+} at 322.631 56.2372
\put {+} at 330.015 56.325
\put {+} at 330.845 59.1456
\put {+} at 321.566 59.0411
\put {+} at 321.297 57.5889
\put {+} at 321.776 56.8367
\put {+} at 330.463 56.5911
\put {+} at 321.208 58.4289
\put {+} at 321.173 58.2356
\put {+} at 330.993 59.0572
\put {+} at 330.148 56.3294
\put {+} at 321.879 59.4525
\put {+} at 330.089 56.2867
\put {+} at 322.354 56.2911
\put {+} at 321.315 57.3206
\put {+} at 321.74 59.4258
\put {+} at 330.398 56.3892
\put {+} at 321.634 56.7917
\put {+} at 331.188 57.1667
\put {+} at 330.01 56.1206
\put {+} at 330.954 56.8431
\put {+} at 321.706 59.4819
\put {+} at 331.33 57.3125
\put {+} at 331.406 57.4639
\put {+} at 331.539 57.9403
\put {+} at 320.942 58.2578
\put {+} at 321.04 57.5469
\put {+} at 321.046 58.7608
\put {+} at 330.048 56.06
\put {+} at 320.988 58.6539
\put {+} at 331.599 57.7164
\put {+} at 320.836 58.1758
\put {+} at 320.988 58.7942
\put {+} at 321.182 59.1319
\put {+} at 320.831 57.9025
\put {+} at 322.257 56.0772
\put {+} at 320.773 57.9658
\put {+} at 320.799 58.4653
\put {+} at 321.072 57.0983
\put {+} at 330.408 56.1481
\put {+} at 320.911 58.8878
\put {+} at 320.83 57.5097
\put {+} at 320.77 57.6703
\put {+} at 330.544 56.2044
\put {+} at 330.947 56.5086
\put {+} at 331.808 58.3089
\put {+} at 331.859 58.1367
\put {+} at 322.017 56.0894
\put {+} at 331.9 58.31
\put {+} at 321.445 56.5044
\put {+} at 331.867 58.5378
\put {+} at 320.731 57.3825
\put {+} at 320.709 58.8406
\put {+} at 320.973 59.2581
\put {+} at 320.54 57.9958
\put {+} at 320.537 57.9158
\put {+} at 320.535 58.3811
\put {+} at 320.789 59.0461
\put {+} at 320.789 59.0706
\put {+} at 320.577 58.6394
\put {+} at 330.612 56.0858
\put {+} at 320.586 57.5725
\put {+} at 332.002 57.9494
\put {+} at 331.957 58.5517
\put {+} at 320.703 58.9517
\put {+} at 320.909 59.2706
\put {+} at 332.005 58.4428
\put {+} at 332.045 57.9869
\put {+} at 320.856 56.9831
\put {+} at 320.969 56.8364
\put {+} at 331.997 58.6411
\put {+} at 331.984 58.6978
\put {+} at 332.099 58.1889
\put {+} at 331.984 57.5103
\put {+} at 320.49 58.8128
\put {+} at 331.963 57.3342
\put {+} at 331.995 58.91
\put {+} at 332.149 57.7558
\put {+} at 320.875 59.4911
\put {+} at 332.19 58.6264
\put {+} at 332.174 58.7872
\put {+} at 332.226 58.6394
\put {+} at 332.146 58.9228
\put {+} at 332.345 58.295
\put {+} at 320.508 57.0153
\put {+} at 320.25 58.8533
\put {+} at 320.1 58.2517
\put {+} at 320.88 56.515
\put {+} at 332.405 57.8289
\put {+} at 332.368 57.6408
\put {+} at 320.088 58.5678
\put {+} at 320.079 58.5294
\put {+} at 320.139 58.7519
\put {+} at 332.47 58.2142
\put {+} at 331.613 56.4447
\put {+} at 320.209 59.025
\put {+} at 332.515 58.2567
\put {+} at 332.102 59.3458
\put {+} at 331.573 56.3086
\put {+} at 319.981 57.4733
\put {+} at 332.645 58.3258
\put {+} at 320.622 56.5092
\put {+} at 331.885 56.4858
\put {+} at 332.618 57.6669
\put {+} at 332.541 58.8722
\put {+} at 319.847 58.5486
\put {+} at 332.66 57.8025
\put {+} at 319.823 58.4883
\put {+} at 320.35 56.7422
\put {+} at 319.785 58.2017
\put {+} at 332.367 59.2875
\put {+} at 332.683 58.6233
\put {+} at 319.872 58.8219
\put {+} at 332.454 59.1942
\put {+} at 320.888 56.1533
\put {+} at 332.588 57.2858
\put {+} at 320.45 56.5131
\put {+} at 332.712 57.5
\put {+} at 320.712 56.2214
\put {+} at 319.956 59.2375
\put {+} at 332.866 58.3533
\put {+} at 319.736 58.8264
\put {+} at 332.87 58.4619
\put {+} at 319.98 56.9844
\put {+} at 319.658 57.6656
\put {+} at 331.711 56.0792
\put {+} at 319.568 58.3533
\put {+} at 320.859 56.0008
\put {+} at 332.963 58.1794
\put {+} at 319.544 58.0011
\put {+} at 319.991 59.4394
\put {+} at 332.262 56.5417
\put {+} at 319.517 58.3867
\put {+} at 332.097 56.3469
\put {+} at 319.772 57.1022
\put {+} at 332.735 57.0933
\put {+} at 319.693 59.2561
\put {+} at 331.935 56.0567
\put {+} at 320.005 56.5839
\put {+} at 320.278 56.2908
\put {+} at 319.58 59.1536
\put {+} at 319.716 59.3806
\put {+} at 319.512 59.0831
\put {+} at 332.757 56.8578
\put {+} at 332.347 56.365
\put {+} at 319.653 59.3619
\put {+} at 319.616 57.0381
\put {+} at 320.324 56.1861
\put {+} at 319.513 57.1878
\put {+} at 319.653 56.9414
\put {+} at 319.622 59.4147
\put {+} at 332.192 56.1103
\put {+} at 319.909 56.4811
\put {+} at 319.547 56.8392
\put {+} at 319.644 56.5758
\put {+} at 319.814 56.3808
\put {+} at 332.513 56.2053
\put {+} at 319.917 56.1967
\put {+} at 319.561 56.5125
\put {+} at 332.523 56.0114
}
\put {\large 22.1} at 321.0 55.85
\put {\large 22.0} at 322.5 55.85
\put {\large 21.9} at 324.0 55.85
\put {\large 21.8} at 325.5 55.85
\put {\large 21.7} at 327.0 55.85
\put {\large 21.6} at 328.5 55.85
\put {\large 21.5} at 330.0 55.85
\put {\large 21.4} at 331.5 55.85
\put {\large R.A. (J2000)} at 326.5 55.65
\put {\large \begin{sideways}56\end{sideways}} at 319.2 56
\put {\large \begin{sideways}57\end{sideways}} at 319.2 57
\put {\large \begin{sideways}58\end{sideways}} at 319.2 58
\put {\large \begin{sideways}59\end{sideways}} at 319.2 59
\put {\large \begin{sideways}Dec (J2000)\end{sideways}} at 318.8 57.5
\endpicture

\caption{\label{obsfield} IC\,1396 field investigated in this paper. The
background image is the extinction map obtained by accumulated star counts
using the 2MASS J-band data (see Sect.\,\ref{extmaps}). Black represents
regions with high extinction, gray zero extinction, and white negative
extinction (caused by star clusters). The small squares mark the fields
observed with MAGIC in JHK and in the 1-0\,S(1) line of H$_2$ (IC\,1396\,W, the
western most field, is marked also, since it is included in the discussion).
Circles correspond to the globules listed in the bottom half of
Table\,\ref{globules} (a few globules are slightly outside the region shown
here). The numbers at the side of circles and small squares are identical to
the globule numbers used in Table\,\ref{globules}. The area marked with dashed
lines was observed with  the Schmidt telescope in Tautenburg in I and [SII].
The very dark, circular region  at about (21$^{\rm h}\!\!$.73:58.8$\degr$) is
not a globule but the very bright star \object{$\mu$\,Cep}, which caused 
non-detections of stars in its vicinity. A similar fake globule is generated by
the  star \object{$\zeta$\,Cep} at (22$^{\rm h}\!\!$.19:58.2$\degr$). The +
signs mark all IRAS sources in the field.} 
\end{figure*}

The weather conditions during the observations of the standard stars were 
photometric. Our ten globules in IC\,1396 were also mainly observed under
photometric conditions. For each globule, we obtained at least one full set of
mosaics (JHK), and we took care to assure that at least one of these sets was
observed under photometric conditions, so that a calibration of the whole
dataset is possible. In case of cirrus, some of the  globules were observed
longer to obtain a uniform limiting magnitude in all fields. The narrow-band
observations of the globules in the H$_2$-filter were done mostly under
non-photometric conditions or at higher air mass. The seeing conditions during
our run were excellent (below 1\arcsec), but we are limited by the large pixel
scale of 1\arcsec$\!\!$.6 per pixel. All broad-band images, IC\,1396 globules
as well as standard stars, were obtained in the airmass range between 1.07 and
1.5. The positions of the observed globules are shown in Fig.\,\ref{obsfield}
(small squares). 

Our standard NIR data reduction included flat-fielding (using sky-flats),
sky-subtraction, and co-addition to mosaics. To co-center the images into a
mosaic we used all stars in the field to ensure a high astrometric accuracy.
Mosaicing and sky-subtraction was done using the xdimsum package in
IRAF\footnote{IRAF is distributed by the National Optical Astronomy
Observatories, which are operated by the Association of Universities for
Research in Astronomy, Inc., under cooperative agreement with the National
Science Foundation.} (Stanford et al. \cite{1995ApJ...450..512S}). For each
filter and globule we produced mosaics of the 3x3 dither patterns.
Additionally, all single mosaics of each filter were stacked to a deep mosaic.
A plate solution for each mosaic was obtained using the 2MASS sources in the
field.

\subsection{Optical data}
 
A large-scale optical survey of the IC\,1396 region was carried out with the
Schmidt camera at the 2-m telescope of the Th\"uringer Landessternwarte
Tautenburg in two observing runs  (9.-12. Sept. 1999 and 21.-25. June 2004). We
observed with a narrow-band filter centred on the [SII] emission line as well
as with a broad-band I-filter, to be able to  identify emission regions. These
optical observations cover about 12\,sqdeg. The survey field is shown  in
Fig.\,\ref{obsfield}. A standard data reduction was performed including bias
subtraction  and flat-field correction. Emission features were searched for by
comparing the narrow-band [SII] images with the I-band images.

\section{Photometry}

\label{photometry}

\subsection{Technique}

For source detection, we applied the SExtractor (Bertin \& Arnouts
\cite{1996A&AS..117..393B}) to our deep K-band mosaic of the globule. The
relative offsets between the positions in this 'master' mosaic and all other
mosaics of the same field were determined by measuring the pixel coordinates of
a bright star in each image. Applying these offsets to the source catalogue of
the 'master' mosaic, we obtained catalogues for all mosaics of the field.
Subsequently, we performed aperture photometry for each object in the
catalogues using the daophot package within IRAF. Because of  the large pixel
scale, the seeing is more or less constant in all mosaics. Therefore, we
decided to hold the aperture for the photometry constant for all mosaics. Sky
coordinates for the sources were determined by applying the plate solution to
the pixel coordinates. As the result we obtained for each
globule a database with sky coordinates and instrumental magnitudes in JHK and
H$_2$.

The instrumental magnitudes of the standard stars were measured in a very
similar way as the IC\,1396 object values, i.e. with aperture photometry using
daophot. Since all these stars were observed under photometric conditions and
airmass similar to the IC\,1396 object fields, their colours can be directly
compared with them. According to the spectral types given in the literature,
our standard stars span a spectral range from B0 to K8, hence they can be used
to establish a main sequence in the colour-colour diagram. 

Tests were performed to verify the reliability of our instrumental photometry
(e.g. repeated observations, field overlap). Only in one of the ten fields
(\object{IRAS\,21539+5821}) were larger systematic photometry mismatches found,
and hence the field was excluded from further analysis. For a more detailed
discussion see Appendix\,\ref{reliability}.

\begin{figure}[t]
\centering
\resizebox{\hsize}{!}{\includegraphics[angle=-90,width=6.5cm]{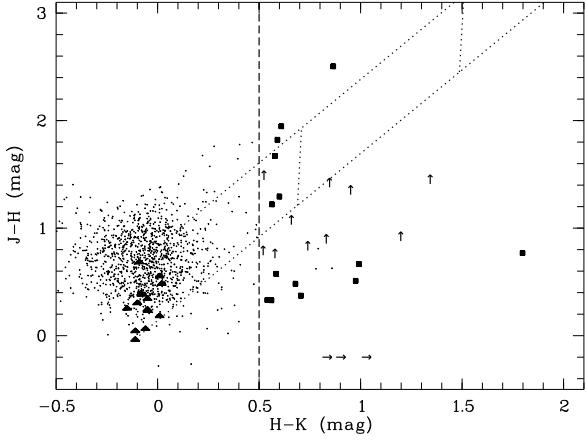}}
\caption{\label{cc1} Colour-colour diagram of the \object{IRAS\,21324+5716}
globule. Small dots are objects in and around the globule. Filled triangles are
field stars with known  spectral type (which mark the unreddened main
sequence). Filled squares, arrows up (lower limit in J-H), and arrows right
(lower limit in H-K and arbitrary J-H) mark reddened objects. The dashed line
shows the criterion used to select these red sources. Dotted lines indicate the
interstellar extinction path according to the extinction law of Mathis
(\cite{1990eism.conf...63M}). The two nearly vertical dotted lines mark
extinctions of $A_{\rm V}$\,=\,10 and 20\,mag. Note that the colour scales are
in instrumental units.}
\end{figure}
 
\subsection{Colour-colour diagrams and reddened sources}
\label{cc} 
 
\begin{figure}[t]
\centering
\resizebox{\hsize}{!}{\includegraphics[angle=-90,width=6.5cm]{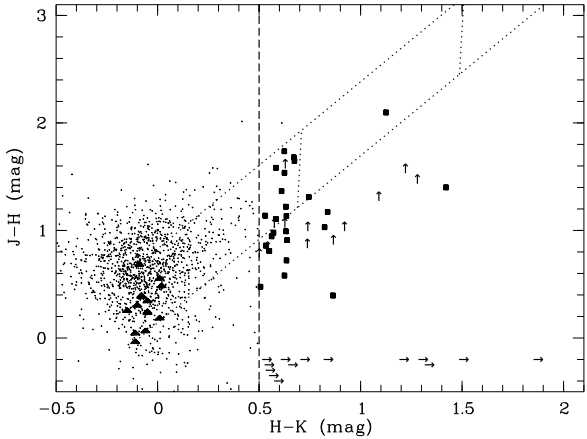}}
\caption{\label{cc2} Colour-colour diagram of the \object{IRAS\,21346+5714}
globule. Symbols as in Fig.\,\ref{cc1}.}
\end{figure}

\begin{figure}[t]
\centering
\resizebox{\hsize}{!}{\includegraphics[angle=-90,width=6.5cm]{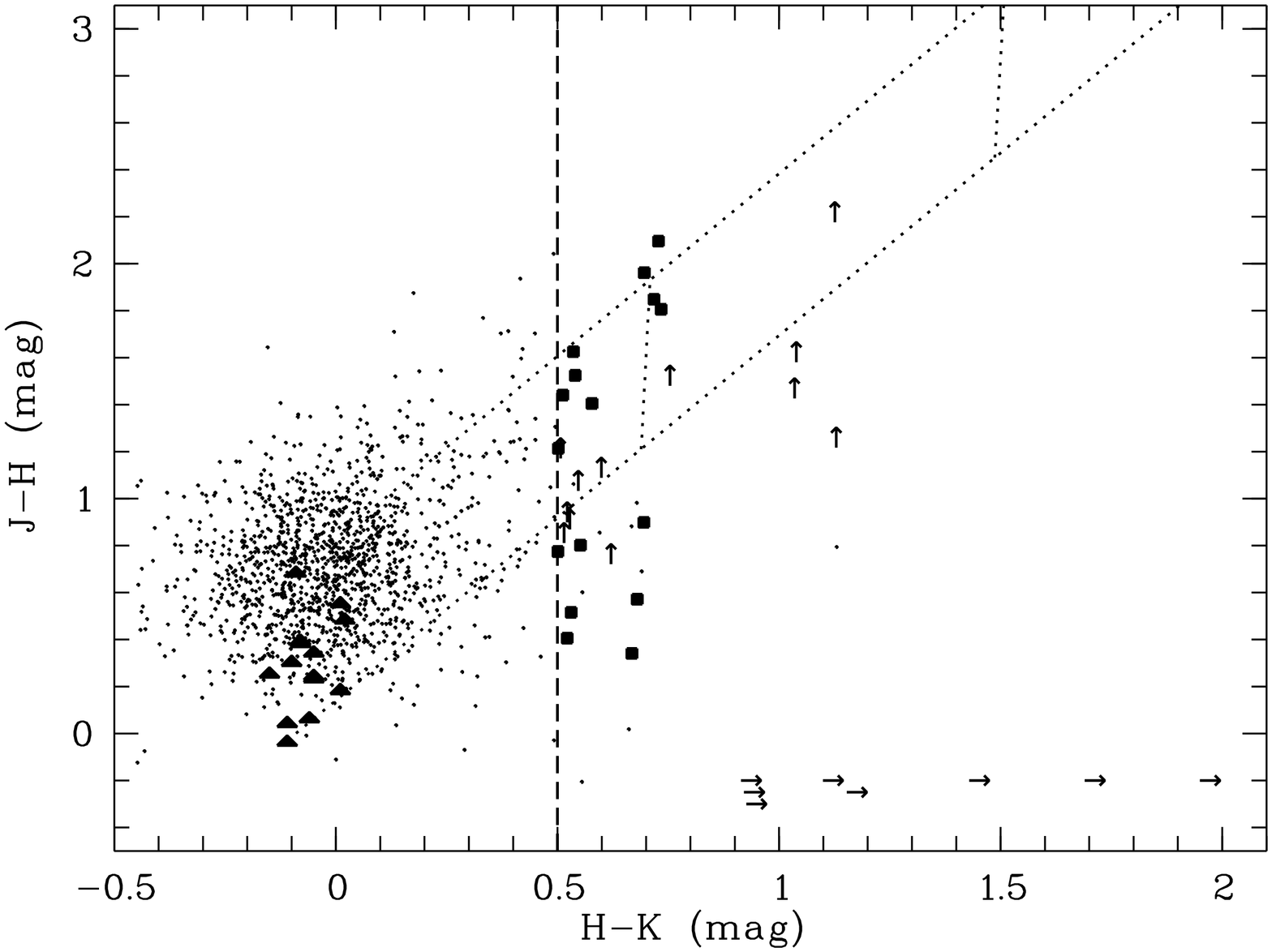}}
\caption{\label{cc3} Colour-colour diagram of the \object{IRAS\,21352+5715}
globule. Symbols as in Fig.\,\ref{cc1}.}
\end{figure}

\begin{figure}[t]
\centering
\resizebox{\hsize}{!}{\includegraphics[angle=-90,width=6.5cm]{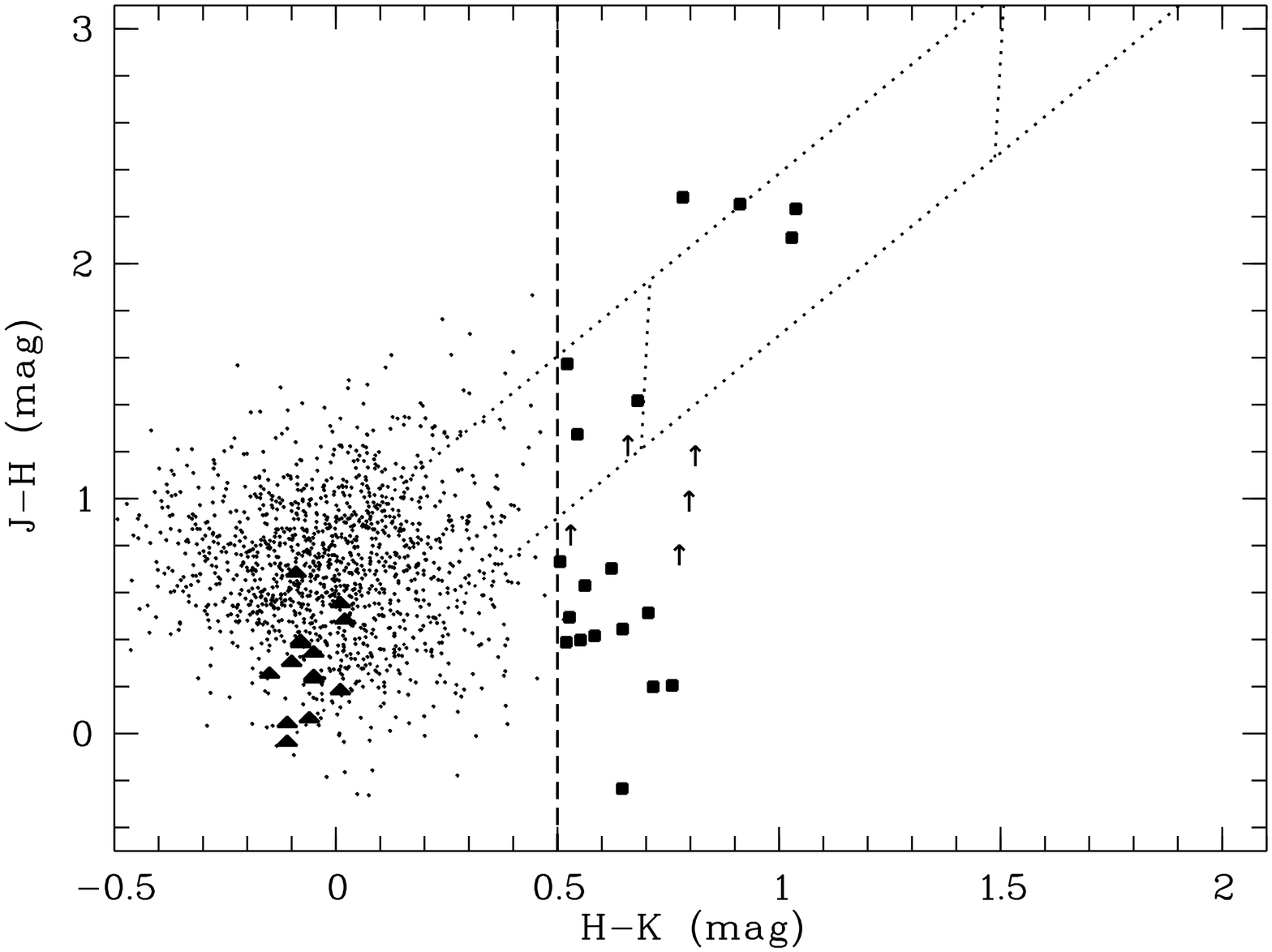}}
\caption{\label{cc4} Colour-colour diagram of the \object{IRAS\,21445+5712}
globule. Symbols as in Fig.\,\ref{cc1}.}
\end{figure}

\begin{figure}[t]
\centering
\resizebox{\hsize}{!}{\includegraphics[angle=-90,width=6.5cm]{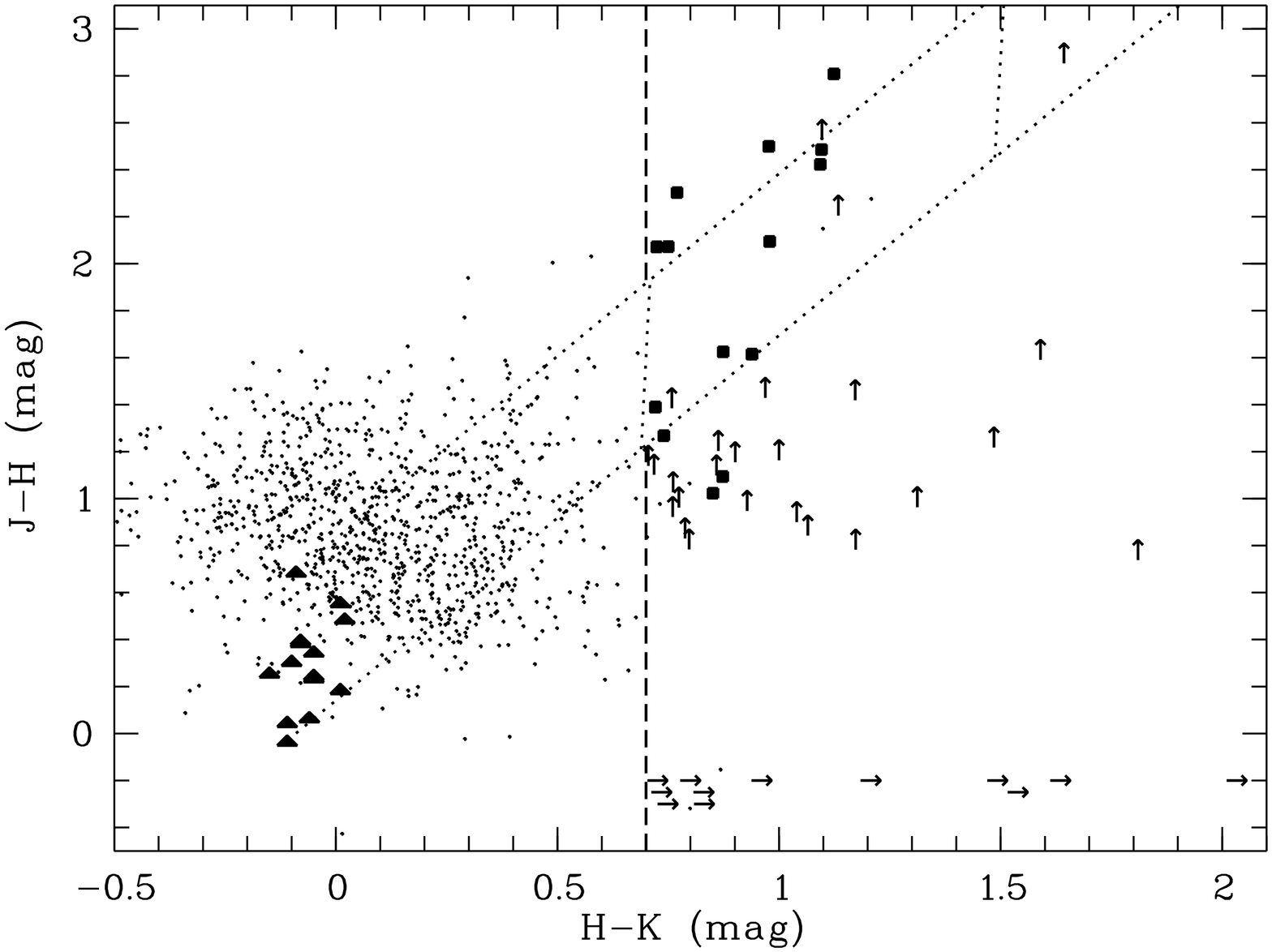}}
\caption{\label{cc5} Colour-colour diagram of the IRAS\,22051+5848 globule.
Symbols as in Fig. \ref{cc1}. For this globule the selection criterion for the
reddened  objects was shifted to H-K\,$>$\,0.7\,mag, because {\it all} objects
in this field show  reddening of 0.2\,mag.}
\end{figure}

We show colour-colour diagrams for five of the nine globules, which contain
more than 20 reddened objects, in Figs.\,\ref{cc1}-\ref{cc5}. These figures
contain a) datapoints for the IC\,1396 objects as small dots, b) the main
sequence datapoints as filled triangles, c) the extinction path calculated with
the extinction law given by Mathis (\cite{1990eism.conf...63M}) as dotted
lines, d) reddened objects as filled squares and arrows. To allow a reliable
comparison between the objects in IC\,1396 and the standard stars, these
diagrams are shown in {\it instrumental} magnitudes, which avoids colour
mismatches. A rough transformation into the 2MASS photometric system can be
obtained by adding about 0.3\,mag to the H-K values. The diagrams show {\it
all} datapoints in the 13\arcmin$\!\!$.5$\times$\,13\arcmin$\!\!$.5 field for
each globule. We generally selected objects with H-K\,$>$\,0.5\,mag as reddened
sources. A more detailed discussion of the colour-colour diagrams and
transformation of instrumental to 2MASS colours can be found in
Appendix\,\ref{cc_cal}.

As noted above, we found a large population of reddened objects for  five of
the nine globules. The remaining four globules contain only a small  number of
reddened sources. To assess the nature of all these red objects, we  estimated
the number of field stars in the direction of IC\,1396 using the  Besan\c{c}on
Galaxy model of Robin et al. (\cite{2003A&A...409..523R}). This  model is
available  online\footnote{{\it 
http://www.obs-besancon.fr/www/modele/modele\_ang.html}} and is able to
generate  a photometry catalogue of objects in a certain Galactic direction
with user-defined  colour constraints. Our selection criterion for red objects
was H-K\,$>$\,0.5\,mag in instrumental colours, which translates into about
H-K\,$>$\,0.8\,mag in absolute colours. 

In a first step, we executed several simulations with standard Galactic optical
extinction (0.7\,mag/kpc) and found that the number of objects with
H-K\,$>$\,0.8\,mag (and even with H-K\,$>$\,0.5\,mag) is zero. This Galaxy
model, however, does not include ultracool dwarfs  with spectral types L and T.
According to the DUSTY models of Chabrier et al. (\cite{2000ApJ...542..464C}),
evolved ultracool objects with T$_{\rm eff}$\,=\,1700\,K have an H-K colour
above 0.8\,mag, and thus could be selected as red objects with our diagrams.
On the other hand, objects with T$_{\rm eff}$ below 1300\,K are too faint to be
seen in our images. For this range of effective temperatures, Gizis et al.
(\cite{2000AJ....120.1085G}) give a space density of $2.11 \cdot
10^{-3}\,pc^{-3}$. Our limiting magnitude constrains the distance at which we
could detect these ultracool objects to $<$\,50\,pc. With these values, the
number of L-type objects in a 1\,sqare degree field is $\approx$\,0.03. From
this estimate, we conclude that the number of Galactic L-type stars in
our red sample is negligible.

For this first calculation, however, we made the assumption of uniform Galactic
extinction, which is not given in our fields, since we observe regions with
high extinction in the direction of IC\,1396. The Besan\c{c}on Galaxy model is
able to include clouds with a given distance and extinction. In a second set of
simulations, we therefore assumed a cloud at a distance of IC\,1396 (750\,pc)
with various extinction values ranging from A$_{\rm V}$\,=\,3\,mag to A$_{\rm
V}$\,=\,10\,mag. The basic effect of such an additional cloud is that the main
sequence in the colour-colour diagram is split. While the foreground objects
remain unreddened, the background stars are shifted along the reddening path.
From these simulations, we found that the background stars begin to exceed the
H-K\,=\,0.8\,mag limit if the extinction of the cloud is higher than A$_{\rm
V}$\,=\,5\,mag. Thus, we expect a significant number of reddened background
stars for globules with A$_{\rm V}$\,$>$\,5\,mag. According to the Galaxy
model, these reddened objects will be mostly M type giants for a cloud with
A$_{\rm V}$\,=\,5\,mag, but with stronger extinction stars with earlier
spectral types also would appear to have H-K\,$>$\,0.8\,mag.

Based on these results, we can make the following conclusions about the nature
of the reddened objects: We can exclude that our reddened objects contain a
significant fraction of foreground stars. This includes also ultracool dwarfs
with spectral types L and T. For globules with high extinction, we expect a
significant number of highly reddened background stars, mostly late-type
giants, along the reddening path. The remaining reddened sources, particularly
the objects whose colours place them below the reddening path in the
colour-colour diagram, should be young stellar objects (YSO) associated with
the globules in IC\,1396, which are intrinsically red because of excess
emission from circumstellar matter. From photometry alone, it is not possible
to unambiguously distinguish between YSOs and background stars.

Five of our nine globules clearly show a cumulation of reddened objects,
whereas the number of reddened objects in the remaining four globules is $<$10.
It is now interesting to analyse whether an excessive number of red objects is
caused by a high  density of YSOs in the globule or a high cloud extinction,
leading to a high number  of background stars which appear reddened. From the
position of the reddened objects in the colour-colour diagram, we are able to
get a rough idea of whether they are  mostly YSOs or background stars, since
background stars will be reddened along the reddening path, whereas YSOs may
also appear below the extinction path. We define as YSO candidates all objects
whose photometry does not rule out the possibility that they are located below
the reddening band. This includes all objects with photometry in all three
bands, which are below the reddening band, but also objects with upper limit in
J, that could be located below the extinction band, and all reddened objects
for which only K-band magnitudes are available. For each globule, we counted
the number of YSO candidates and the total number of reddened objects. For
globules without a large reddened population, it was found  that only 77\,\% of
the red sources are YSO candidates, whereas the remaining objects are in the
extinction path. On the other hand, for globules with a large  reddened 
population, the fraction of YSO candidates is 68\%, i.e. significantly lower
than in the 'empty' globules. The most straightforward interpretation of  this
result is that the globules with many red objects have both a high density  of
YSO and strong extinction, leading to a high number of reddened background
stars.  The other globules possess smaller extinction values and fewer young
stellar objects.

Assuming a distance of 750\,pc for IC\,1396 and no significant extinction, our
reddened targets have absolute J-band magnitudes between 3.1 and 7.3\,mag. This
constrains the masses for these objects to a range roughly between 0.05 and 
0.9\,M$_{\odot}$ (Baraffe et al. \cite{1998A&A...337..403B}). Note that there
are also no brighter sources significantly below the reddening path but with
H-K\,$<$\,0.5\,mag. Hence no unextincted Herbig AeBe stars are present. Since
the globules show J-band extinction of at least 0.9\,mag (see
Table\,\ref{globules}), there could be a few higher mass stars in our sample.
On the other hand, it is unlikely that our faintest objects have substellar
masses, rather than just being strongly influenced by extinction.

\section{Extinction maps}
\label{extmaps}
 
\subsection{Method}
 
We determined extinction maps of the region 21$^{\rm h}\!\!$.3 to 22$^{\rm
h}\!\!$.2 in R.A. and 56$^\circ$ to 60$^\circ$ in DEC (J2000) using accumulated
star counts (Wolf-diagrams) in the 2MASS database. An example Wolf-diagram is
shown in Fig.\,\ref{wolfdiag}. Stars are counted in
3\arcmin\,$\times$\,3\arcmin\, boxes every 20\arcsec\, down to the
completness limit of the catalogue. A co-centred 1$^\circ$\,x\,1$^\circ$ sized
field was chosen as "unextincted" comparison. Minimum and maximum traceable
extinction values are calculated and shown in Fig.\,\ref{ranges}, depending on
the resolution used. We can trace about one to 20\,mag optical extinction.
Details of the star count method and the determination of the detection limits
can be found in Appendix\,\ref{exmade}.

\subsection{Results}

\begin{figure*}[t]
\centering
\includegraphics[angle=-90,width=8.5cm]{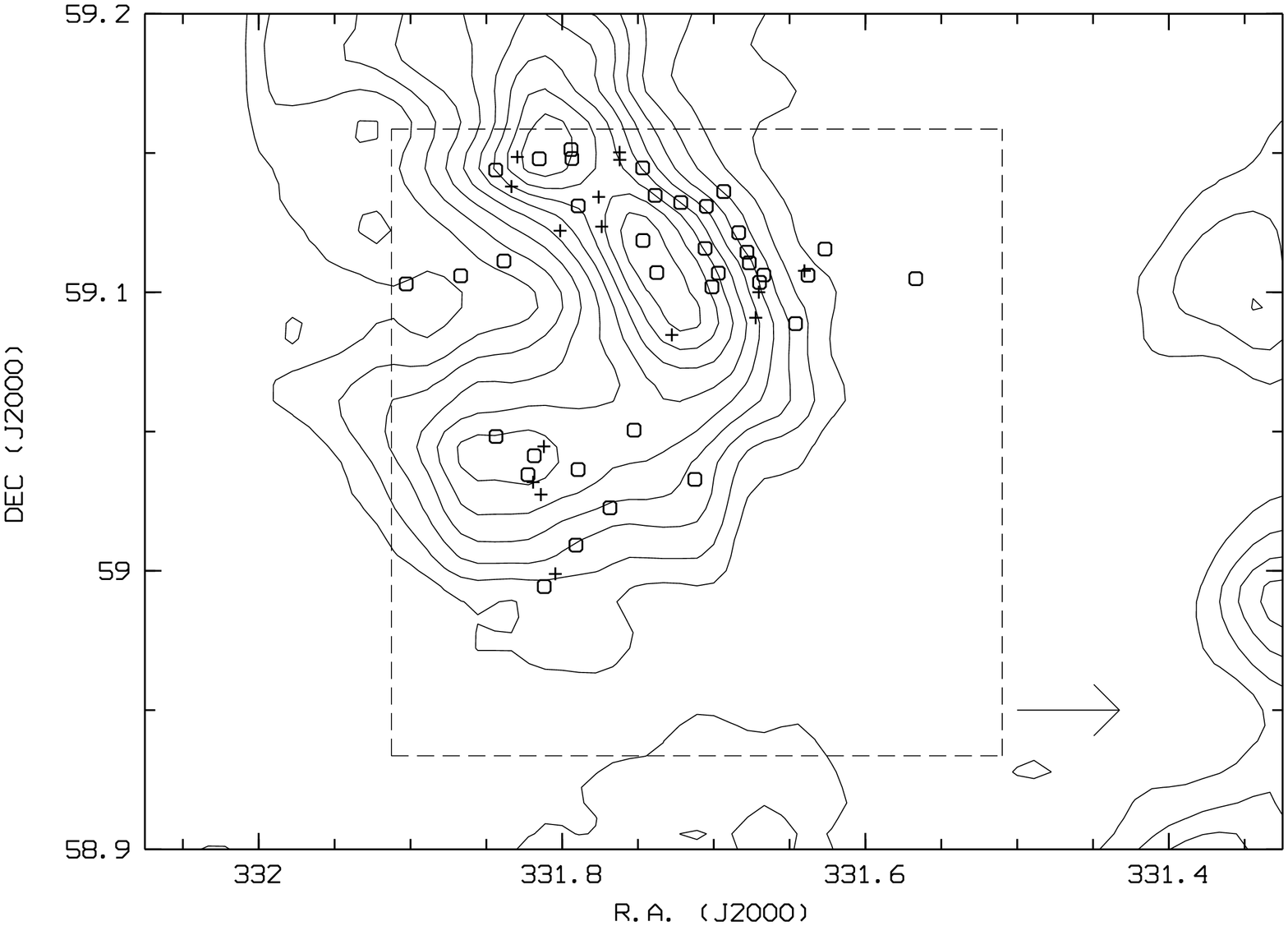}\hfill
\includegraphics[angle=-90,width=8.5cm]{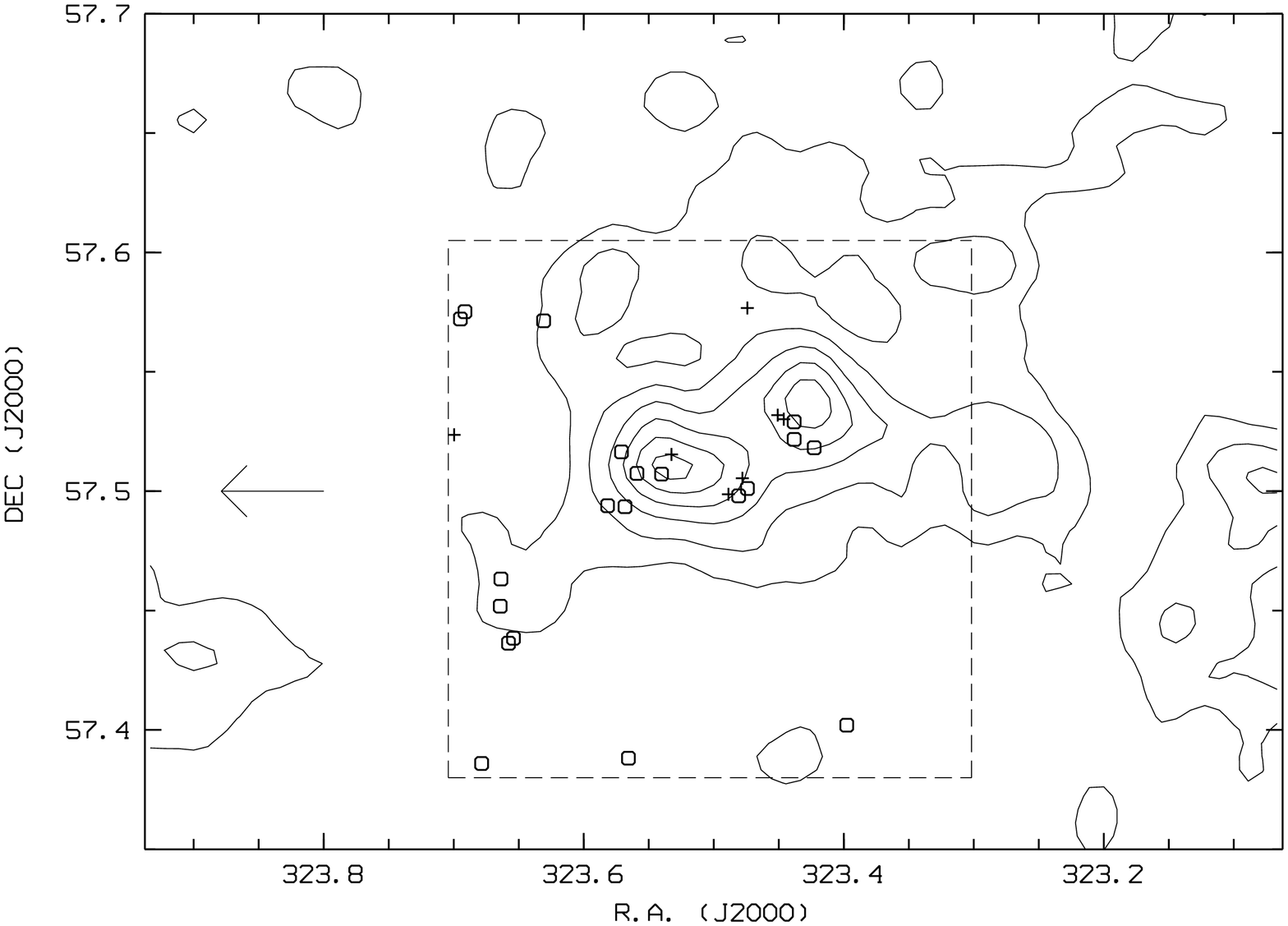}\\
\includegraphics[angle=-90,width=8.5cm]{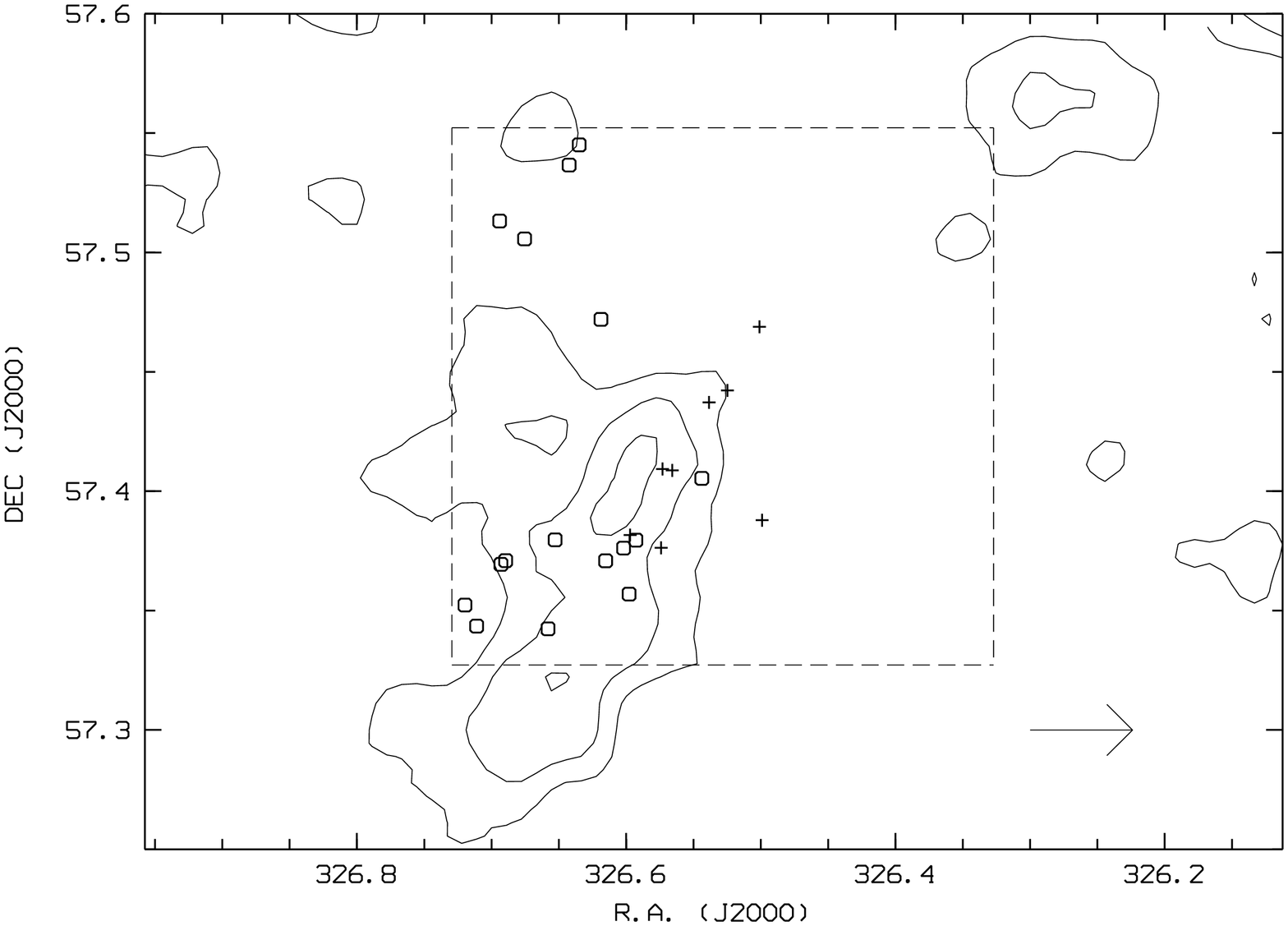}\hfill
\includegraphics[angle=-90,width=8.5cm]{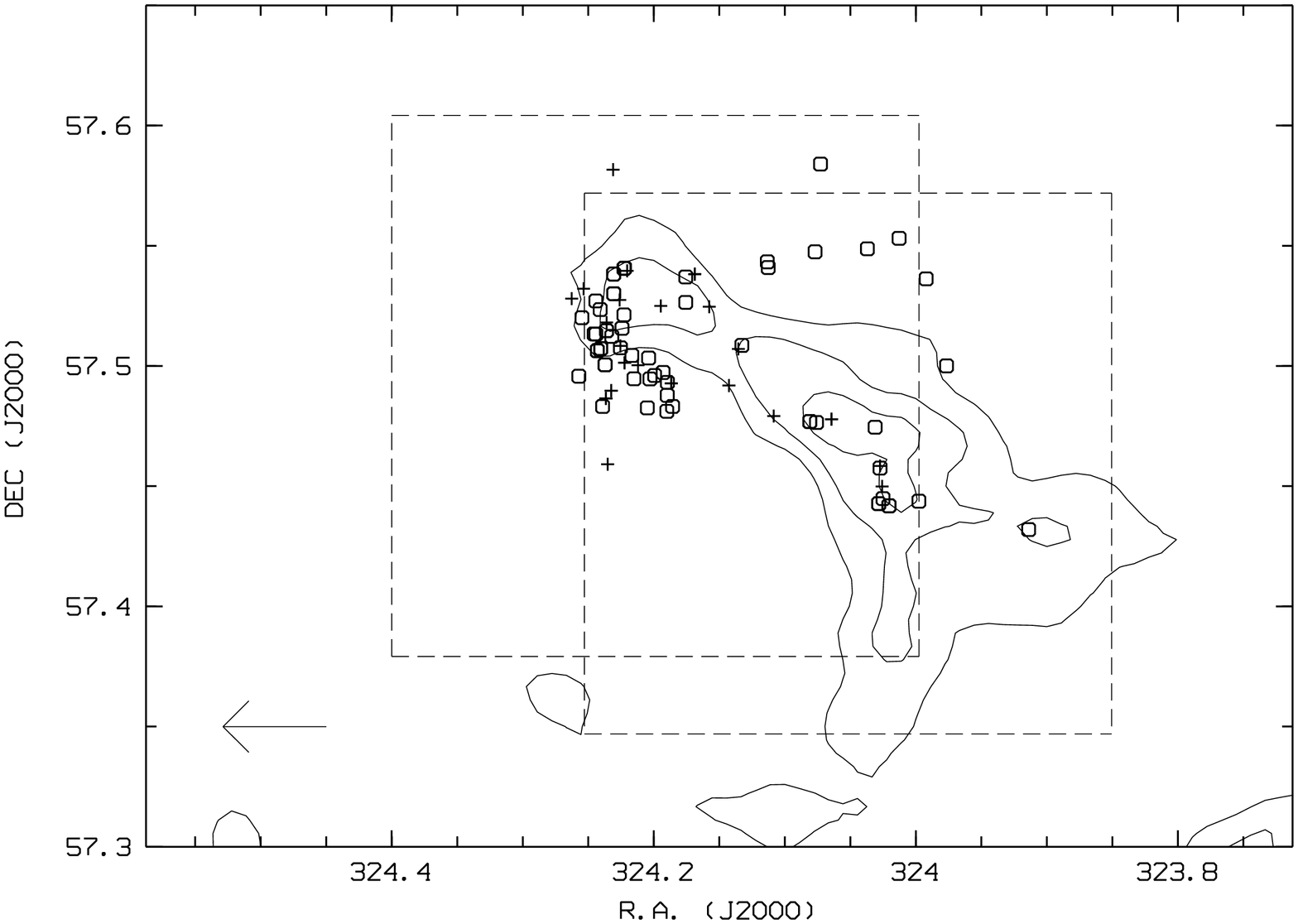}\\
\caption{\label{cont_glob} Contour plots of the extinction obtained from
the J-band 2MASS data for the globules 11 (upper left), 3 (upper right), 9
(lower left), and 4 and 5 (lower right). Contours start at 1\,mag optical
extinction and increase by 1\,mag. Conversion from extinction in the J-band to
optical extinction was done according to Mathis et al.
(\cite{1990eism.conf...63M}). Dashed lines mark the borders of the observed NIR
fields. $+$ signs indicate reddened objects in the reddening path, while
circles mark the YSO candidates, and arrows the principle direction towards the
O6.5V star HD\,206267. Note that the contours are slightly smoothed and hence
peak extinction values are somewhat lower than in Table\,\ref{globules}.} 
\end{figure*}

The extinction map of the whole IC\,1396 field shows several regions of high
extinction (see Fig.\,\ref{obsfield}). These are the globules around the star
HD\,206267, and a further group towards the north-east. We defined a 'globule'
as an object with an extinction of at least 3\,$\sigma$ above the noise level
and a size of at least 9 square arcminutes. With these criteria, an automated
search using the SExtractor (Bertin \& Arnouts  \cite{1996A&AS..117..393B})
revealed 20 globules in IC\,1396. Five of these globules were contained in the
target list for our near-infrared survey with MAGIC from Schwartz et al.
(\cite{1991ApJ...370..263S}). The globules in this list that could not be
detected in our extinction maps possess sizes and masses below our detection
limits (see also Patel et al. \cite{1995ApJ...447..721P}). Adding the remaining
five objects from this list, there are 25 dark  clouds in the field. The large
region of high extinction at about (21$^{\rm h}\!\!$.73:58.8$\degr$) is caused 
by the bright star $\mu$\,Cep, which prevented the detection of stars in its
vicinity. Similarly the star $\zeta$\,Cep at (22$^{\rm h}\!\!$.19:58.2$\degr$)
causes such a fake globule. The full list of all globules with their
coordinates and known identifications is given in Table\,\ref{globules}. 
Figure\,\ref{cont_glob} shows as examples the extinction maps for the globules
where we detected a large number of reddened sources.

To measure the mass we integrated the total extinction in the globules (A$_{\rm
V}^{\rm tot}$). Only regions with an extinction above the one sigma noise level
where considered, and the outer radius of the globule was taken as the size
given by the SExtractor software (FWHM, assuming a Gaussian core). From these
integrated extinction values we determined the mass of the globules using the
fact that the column density N(H) of hydrogen atoms can be expressed as
6.83\,10$^{21}$\,cm$^{-2}$\,$\times$\,A$_{\rm V}$/R$_{\rm V}$ where R$_{\rm V}$
is typically 3.0 (see e.g. Mathis \cite{1990eism.conf...63M}). Using a distance
of 750\,pc for the globules, and the 20\arcsec\, pixel scale in our maps, the
total mass in a globule can be determined from the integrated optical
extinction (A$_{\rm V}^{\rm tot}$) in the globule by: M$_{\rm
glob}$\,$[$M$_\odot]$\,=\,0.098\,A$_{\rm V}^{\rm tot}$\,$[{\rm mag}]$. Note
that the obtained extinction values are averaged over the box-size of
3\arcmin\,$\times$\,3\arcmin. If the dust is concentrated in a smaller area,
the obtained masses are lower limits. The error of the estimated mass depends
on the mean extinction within the globule compared to the noise level in our
map. As a typical value we find that the mean extinction is about 5\,$\sigma$
of the noise level and hence the uncertainties in the inferred masses are about
20\,\%. In the case of IC\,1396\,N we can compare our mass estimate with
literature values. Wilking et al. (\cite{1993AJ....106..250W}) estimate
380\,M$_\odot$ and Serabyn et al. (\cite{1993ApJ...404..247S}) give
480$\pm$120\,M$_\odot$ within the central 0.3\,pc. Given the large
uncertainties and the fact that the object is rather small and the mass is
concentrated close to the actual star forming core (see e.g. Codella et al.
\cite{2001A&A...376..271C}), our lower limit for the whole globule of
300\,M$_\odot$ agrees well with these estimates. Weikard et al.
(\cite{1996A&A...309..581W}) give an estimate of 12000\,M$_\odot$ for the total
mass of their mapped region (about six square degrees centred on HD\,206267).
The mass estimated from our extinction maps in the same field is
9000\,M$_\odot$, reasonably close.

Within the errors of about 20\,\% the total optical extinction values and
globule masses, obtained from the J- and H-band 2MASS data, are consistent. 
Some of the masses calculated with the K-band data differ by a larger amount.
There are two groups of objects: 1) The mass from the K-band data is much
smaller than the mass obtained from the J- and H-band data. The reason could be
that the globule contains an embedded cluster of stars. These stars are easier
to detect in K and hence decrease the apparent extinction. In these cases the
given masses M$_{\rm J,H}$ are lower limits for the globule mass. 2) The
mass from the K-band data is much larger than the mass obtained from the J- and
H-band data. An explanation could be that the conversion of extinction in the
K-band to optical extinction cannot be performed following Mathis et al.
(\cite{1990eism.conf...63M}) that assume an opacity index of 1.7. A lower value
might be valid in these globules. Note that this effect could counter-balance
1) in the case of an embedded cluster.

\section{Globule properties}

\label{activity}
 
\renewcommand{\arraystretch}{0.85}
\begin{table*}[t]
\caption{\label{globules} All detected globules in the IC\,1396 region. The
upper part of the table lists the objects observed with MAGIC in JHK, while in
the lower part the additional globules detected in our extinction maps are
given. Column\,4 lists the projected distance of the globule from the exciting
star HD\,206267. The size in square arcminutes is the value we obtained from
the SExtractor software (a ? marks the globules that were not detected in the
extinction maps.). Column 6 contains the number of reddened objects (in 
brackets: number of YSO candidates as defined in Sect.\,\ref{cc}). In 
columns 7 to 9 we give the peak extinction values in the three filters obtained 
from our extinction maps, while columns 10 to 12 list the masses estimated for 
the globules. $^*$ The given masses are measured for both globules together.} 
\centering
\begin{tabular}{ll|ccccc|rrr|rrr}
Nr. & Name(s) & \multicolumn{2}{|c}{($\alpha$;$\delta$)\,(J2000)} &
Distance & Size & Red stars & A$_{\rm J}$ & A$_{\rm H}$ & A$_{\rm K}$ & M$_{\rm
J}$ & M$_{\rm H}$ & M$_{\rm K}$ \\   
 & & [h\,m] & [\degr\,\arcmin\,] & [pc] & [$\Box\arcmin$] & &
 \multicolumn{3}{|c|}{[mag]} & \multicolumn{3}{|c}{[M$_\odot$]} \\  
\noalign{\smallskip}
\hline
\noalign{\smallskip}
1 & \object{IRAS\,21246+5743} & 21 26 & 57 58 & 24.7 & 114 & 31(18) & 3.1 & 1.9 & 1.3 & 515 & 541 & 401 \\     
 & IC\,1396\,W &  &  &  &  &  &  &  &  &  &  \\                                                  
2 & \object{IRAS\,21312+5736} & 21 33 & 57 50 & 11.7 &?&  9(7) & 0.9 & 0.7 & 0.7 &  \\                    
3 & IRAS\,21324+5716 & 21 34 & 57 32 &  8.6 &  50 & 27(20) & 2.0 & 1.2 & 1.1 & 224 & 182 & 207 \\     
 & \object{LDN\,1093} &  &  &  &  &  &  &  &  &  &  \\                                                    
 & \object{LDN\,1098} &  &  &  &  &  &  &  &  &  &  \\                                             
4 & IRAS\,21346+5714$^*$ & 21 36 & 57 28 &  5.4 &?& 51(38) & 1.1 & 1.0 & 0.7 & 120 & 119 &  74 \\     
5 & IRAS\,21352+5715$^*$ & 21 37 & 57 30 &  3.1 &?& 36(22) & 0.8 & 0.7 & 0.6 & 120 & 119 &  74 \\     
 & \object{LDN\,1099} &  &  &  &  &  &  &  &  &  &  \\                                                    
 & \object{LDN\,1105} &  &  &  &  &  &  &  &  &  &  \\ 
6 & \object{IRAS\,21354+5823} & 21 37 & 58 37 & 15.2 &?&  4(3) & 0.9 & 0.8 & 1.1 &  \\                    
 & \object{LDN\,1116} &  &  &  &  &  &  &  &  &  &  \\                                                    
7 & IRAS\,21388+5622 & 21 40 & 56 36 & 12.3 &?&  7(5) & 0.9 & 0.7 & 0.9 &  \\                    
8 & \object{IRAS\,21428+5802} & 21 44 & 58 17 & 13.8 &?& 10(8) & 1.1 & 0.9 & 0.9 &  \\                    
 & \object{LDN\,1130} &  &  &  &  &  &  &  &  &  &  \\                                                    
9 & IRAS\,21445+5712 & 21 46 & 57 26 & 13.2 &?& 24(16) & 0.9 & 0.8 & 0.7 & 120 & 141 & 195 \\     
 & \object{IC\,1396\,E} &  &  &  &  &  &  &  &  &  &  \\                                                  
10 & IRAS\,21539+5821 & 21 55 & 58 35 & 32.9 & 115 & -- & 2.0 & 1.3 & 0.9 & 374 & 351 & 458 \\      
11 & IRAS\,22051+5848 & 22 07 & 59 08 & 55.9 & 226 & 51(36) & 2.7 & 2.1 & 1.5 & 788 & 714 & 847 \\      
 & \object{LDN\,1165} &  &  &  &  &  &  &  &  &  &  \\                                                    
 & \object{LDN\,1164} &  &  &  &  &  &  &  &  &  &  \\ 
\noalign{\smallskip}
\hline                                             
\noalign{\smallskip}
12 &                  & 21 25 & 57 53 & 25.5 &  77 &    & 1.8 & 1.4 & 1.5 & 279 & 285 & 187 \\     
13 &                  & 21 25 & 58 37 & 29.3 &  50 &    & 1.4 & 0.9 & 1.1 & 160 & 158 & 109 \\     
14 & \object{LDN\,1086}        & 21 28 & 57 31 & 19.4 &  50 &    & 1.5 & 1.1 & 1.0 & 273 & 273 & 233 \\     
15 &                  & 21 33 & 59 30 & 29.2 &  39 &    & 1.1 & 1.1 & 1.1 &  86 & 146 & 160 \\     
16 & \object{LDN\,1102}        & 21 33 & 58 09 & 14.0 &  46 &    & 1.4 & 1.1 & 1.2 & 189 & 182 & 206 \\     
17 & \object{LDN\,1088}        & 21 38 & 56 07 & 18.6 &  64 &    & 1.4 & 1.0 & 0.8 & 298 & 299 & 372 \\     
18 & \object{LDN\,1131}        & 21 40 & 59 34 & 28.2 &  45 &    & 1.6 & 1.2 & 1.1 & 265 & 303 & 305 \\     
19 & IC\,1396\,N      & 21 40 & 58 20 & 11.6 &  84 &    & 1.5 & 1.2 & 1.0 & 285 & 294 & 271 \\     
20 & LDN\,1131        & 21 41 & 59 36 & 28.6 &  87 &    & 1.6 & 1.2 & 1.2 & 390 & 353 & 452 \\      
21 & \object{LDN\,1129}        & 21 47 & 57 46 & 14.9 & 147 &    & 1.4 & 1.0 & 0.9 & 385 & 403 & 498 \\      
22 &                  & 21 49 & 56 43 & 21.3 &  70 &    & 1.2 & 0.9 & 0.8 & 207 & 199 & 311 \\      
23 & \object{LDN\,1153}        & 22 01 & 58 54 & 44.7 & 270 &    & 2.3 & 1.7 & 1.2 & 927 & 875 & 933 \\      
24 & \object{LBN 102.84+02.07} & 22 08 & 58 23 & 53.7 &  74 &    & 1.7 & 1.3 & 0.9 & 402 & 382 & 376 \\      
25 & LBN 102.84+02.07 & 22 08 & 58 31 & 54.1 &  36 &    & 1.3 & 1.1 & 0.9 & 201 & 205 & 247 \\      
\end{tabular}
\end{table*}
\renewcommand{\arraystretch}{1}

With the results of Sect. \ref{photometry} and \ref{extmaps}, a homogeneous
sample of globules is now at our disposal (see Table \ref{globules}). To 
evaluate possible trends in this dataset, we selected the following globule 
properties and tested them for correlations: projected distance of the globule
from the exciting O star HD\,206267, projected size of the globule, number of
YSO candidates within the field of the globule, peak  extinction value, total
mass, mean column density within the globule (total mass  divided by the size).
For each pair of these properties, we fitted their relationship  linearly and
calculated the correlation coefficient $r$. Since the value  $t = r
\sqrt{\frac{(N-2)}{1-r^2}}$ ($N$: number of datapoints) follows  Student's
$t$-distribution, this gives us the probability that the two  parameters show
indeed a significant correlation.  In order to perform these tests, we set
globule sizes to 30 square arcminutes and masses to 120\,M$_\odot$, if
they could not be determined. These values are the upper limits from our
detection procedure for globules in the extinction maps. As peak extinction, we
used the values from the J-band, since they have the best signal-to-noise
ratio.

The best correlations are obtained when comparing size and mass, peak
extinction and mass, and size and peak extinction. The false alarm
probabilities (FAP) for linear relationships between these three parameters are
below $10^{-3}$\,\%. These are  expected correlations, since the mass is
proportional to the mean extinction and  the size of the globule. The
correlation of size and peak extinction might be due  to our method that
smooths the extinction over a 3\arcmin\,$\times$\,3\arcmin\, field, naturally
leading to lower peak extinction values for small globules.

\begin{figure}[t]
\centering
\resizebox{\hsize}{!}{\includegraphics[angle=-90,width=6.5cm]{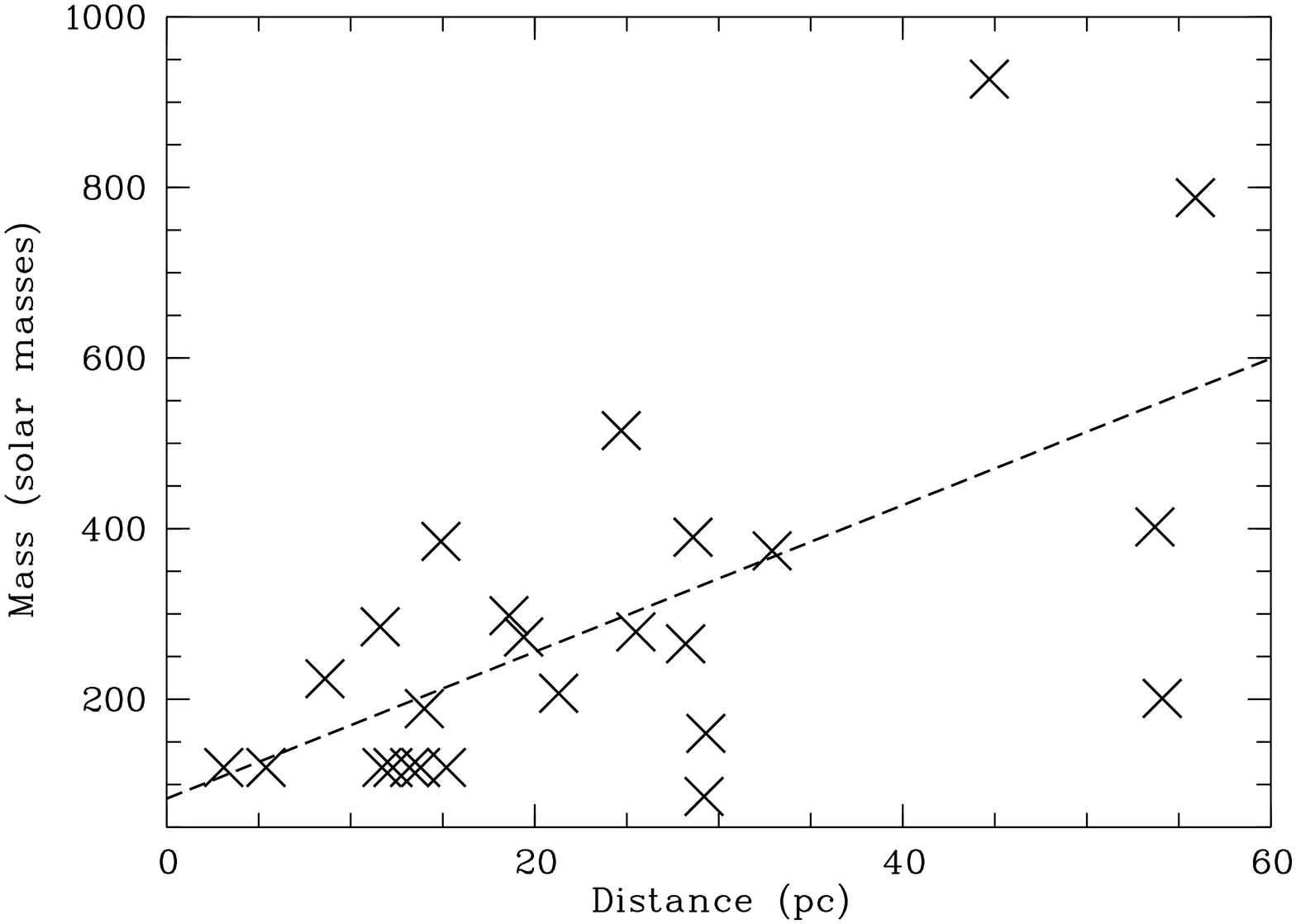}}
\caption{\label{distmass} Globule masses vs. distance from the exciting O star.
A linear least-square fit is shown as dashed line. This correlation has a FAP
of 0.078\,\%. If we exclude the four globules with $d$\,$>$\,40\,pc, the
correlation is still significant with a FAP of 2.2\,\%.}
\end{figure}

Additionally, we found a non-obvious correlation between mass $M$ and distance
$d$ from the exciting star: Globules farther away from HD\,206267 have on
average larger masses. The least-square fit gives $M[M_{\odot}] = 8.6\,d\,[pc]
+ 84$, and the FAP for this correlation is 0.078\,\%.  We repeated this test
excluding the four globules with $d>40$\,pc, because it might be that these
regions are too far away from HD\,206267 to be influenced by its radiation.
Without the data points for these globules, the FAP is still only 2.2\,\%.
Thus, there is a significant connection between globule mass and distance from
the exciting star (see Fig. \ref{distmass}). 

There are two possible explanations for this correlation: 1) The mass loss rate
of a cloud induced by photo-ionisation of the external layers is inversely
proportional to the distance from the exciting star (see e.g Codella et al.
\cite{2001A&A...376..271C}). Hence, more distant globules do not lose as much
mass due to evaporation. 2) Globules closer to the exciting star contain more
young objects  (see below) and hence the mass estimates of these globules lead
to too low values due  to the star count method. Connected to the distance-mass
correlation are correlations of distance and globule size as well as distance
and peak extinction value. These are consequences of the interrelations
described in the above paragraph.

The remaining combinations of parameters show no significant correlations, i.e.
with FAP below 10\,\%. Particularly, we see no correlation between distance and
number of YSO candidates. This can be used to rule out the  second explanation
for the mass-distance correlation. This relationship is therefore most likely
related to mass loss via photo-evaporation, as explained above.  For all
correlation tests that use the number of YSO candidates, we are restricted to
those globules in which red objects were identified (see
Table\,\ref{globules}). Thus, small number statistics hamper the correlation
tests for these parameters. 

\begin{figure}[t]
\centering
\resizebox{\hsize}{!}{\includegraphics[angle=-90,width=6.5cm]{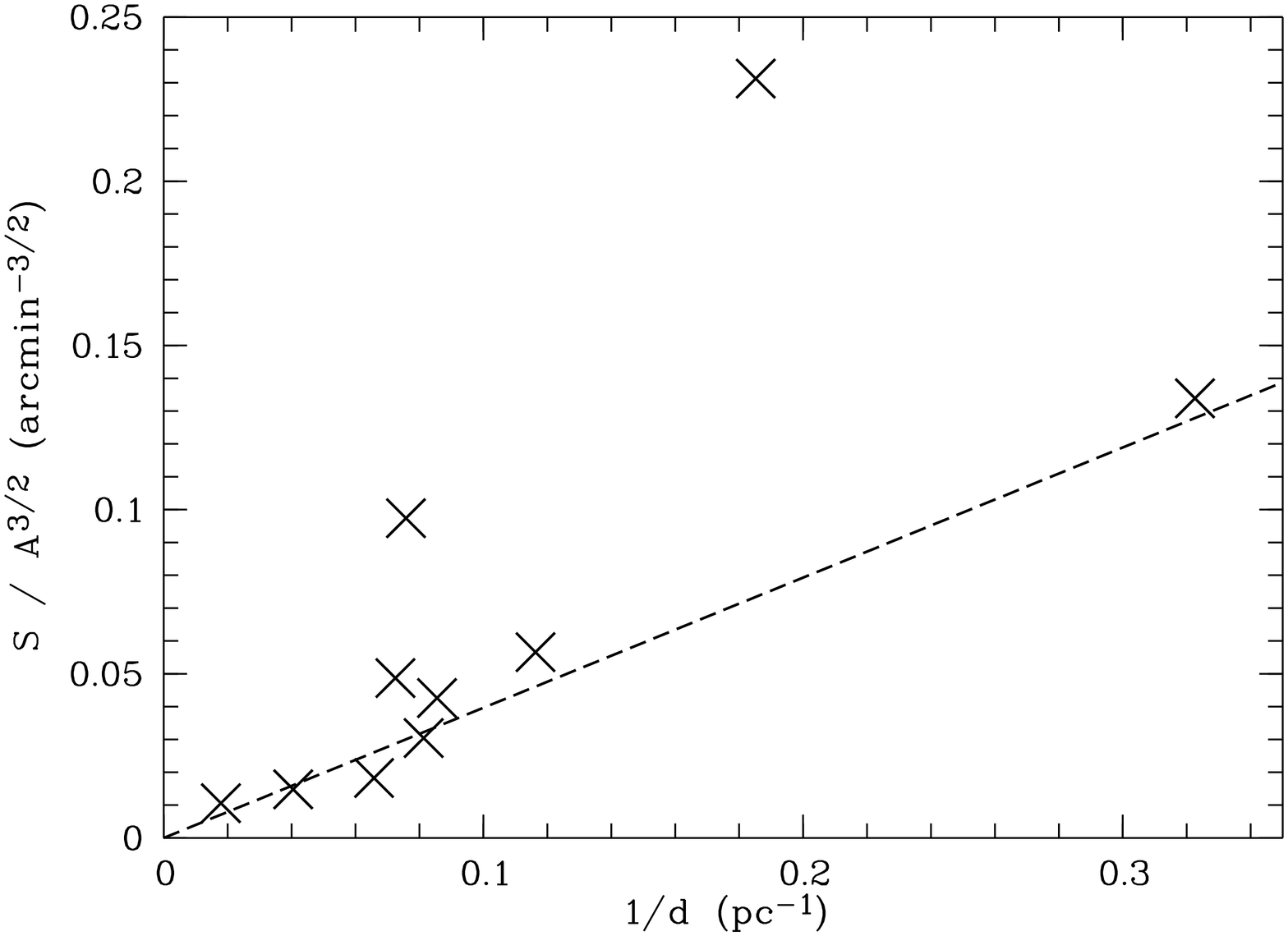}}
\caption{\label{distsa} Diagram of star density vs. inverse distance from the
exciting O star. A linear least square fit is shown as dashed line. The
correlation has a FAP of 2.5\,\%. If we exclude globules with $d$\,$>$\,40\,pc
a FAP of 5.0\,\% is still reached.} 
\end{figure}
 
How do these findings fit in a scenario where the radiation pressure of the O
star is the primary triggering mechanism for star formation in the globules? In a
first order interpretation, we assume that the density of young objects
$S/A^{3/2}$ (S - numer of YSO candidates; A - projected  size of the globule)
is positively correlated with the mass of the globule $M$ and the  pressure $P$
exerted by the radiation flux $\Phi$ from the star: $S/A^{3/2} \propto M \times
P$. The radiation pressure or ionising flux is inversely proportional to the
square of the distance ($P \propto \Phi \propto d^{-2}$; Codella et al.
\cite{2001A&A...376..271C}) and the mass is proportional to the distance ($M
\propto d$ as shown in Fig.\,\ref{distmass}). Hence we should expect a
correlation of the star density with the inverse distance from the star: $S /
A^{3/2} \propto M \times P \propto d^{-1}$. Indeed, such a correlation is seen
tentatively in our data. A linear fit gives  $S / A^{3/2} = 0.555 d^{-1}$,
with a FAP of 2.5\,\% (see Fig. \ref{distsa}). If we exclude the one globule
with $d>40$\,pc, the FAP increases to 5.0\%. Although this result clearly needs
to be substantiated with more datapoints, it tentatively confirms our initial
assumption that the star forming activity is driven by the radiation pressure
of the O star.

We re-examined the conclusion of Sect.\,\ref{cc} where we found that globules
with a rich population of young stars also show high extinction leading to a
high number of reddened background objects. Although there is no correlation
between extinction A$_{\rm J}$ and the number of YSO candidates  $N$, the data
show a clear tendency: All globules with $N$\,$<$\,10 have A$_{\rm
J}$\,$\le$\,1.1\,mag. On the other hand, 50\,\% of the globules with
$N$\,$>$\,10 exhibit A$_{\rm J}$\,$\ge$\,2\,mag. Thus, globules that harbour
many red objects tend to show high extinction values, confirming our result
from Sect.\,\ref{cc}. 

We further investigated the positions of the YSO candidates in the globules
with respect to the direction of the O6.5V star HD\,206267.
Figure\,\ref{cont_glob} shows the positions of the reddened sources overplotted
on the extinction map for the five globules where a large number of reddened
objects was found (the two overlapping fields are combined in one figure). In
three of the fields the YSO candidates (circles) seem to be preferentially
positioned towards the direction of the exciting O6.5V star. Objects within the
reddening path (plus signs) seem to be more concentrated towards the high
extinction regions. This picture agrees well with the expectations of triggered
star formation via radiation-driven implosion, although this behaviour is not
present in all investigated globules.

\section{New HH objects and H$_2$ outflows}

\label{outflows}

\begin{table}
\caption{\label{hhcoord} Coordinates of the brightest knots in the newly
detected Herbig-Haro objects. $^*$ Counterparts of the optical knots detected
in H$_2$ in Froebrich \& Scholz (\cite{2003A&A...407..207F}).}
\centering
\begin{tabular}{lccc}
Object & H$_2^*$ & $\alpha$(J2000) & $\delta$(J2000) \\ 
& & [h\,m\,s] & [\degr\,\,\arcmin\,\,\arcsec\,] \\
\noalign{\smallskip}
\hline
\noalign{\smallskip}
\object{HH\,588\,NE3} & & 21 41 00.0 & 56 37 19 \\
            & & 21 41 01.0 & 56 37 25 \\
\object{HH\,865\,A}  & & 21 44 28.5 & 57 32 01 \\
            & & 21 44 29.3 & 57 32 24 \\
\object{HH\,865\,B}  & & 21 45 10.5 & 57 29 51 \\
\object{HH\,864\,A} & 2-j & 21 26 01.4 & 57 56 09 \\
           & 2-j  & 21 26 02.0 & 57 56 09 \\
\object{HH\,864\,B} & 5 & 21 26 07.9 & 57 56 03 \\
\object{HH\,864\,C} & 2-b & 21 26 21.3 & 57 57 40 \\
          & 2-d & 21 26 18.6 & 57 57 12 \\
\end{tabular}
\end{table}

In our observed field there are several known Herbig-Haro objects and outflows.
These are mainly the well investigated flows in the IC\,1396\,N globule
(\object{HH\,589}-595 found by Ogura et al. \cite{2002AJ....123.2597O} and
\object{HH\,777}-780 discovered by Reipurth et al. \cite{2003ApJ...593L..47R}).
The H$_2$ outflow from IC\,1396\,W which was shown in Froebrich \& Scholz
(\cite{2003A&A...407..207F}) has no published optical emission counterpart.
Connected to IRAS\,21388+5622 is the HH\,588 object (Ogura et al.
\cite{2002AJ....123.2597O}). In the IRAS\,22051+5848 field a giant outflow
\object{HH\,354} was found by Reipurth \& Bally (\cite{2001ARA&A..39..403R}). 

In our [SII] images we were able to re-discover all of the known HH objects
around IC\,1396\,N. The whole field is full of extended emission line objects
(visible as excess emission in the [SII] images compared to the I-band fluxes).
These filaments are excited by the strong UV emission of the O6.5V star
HD\,206267. Many of these filaments show a similar appearance to some of the HH
objects near IC\,1396\,N. Hence, it is very difficult to decide from the shape
of the emission alone if we see an outflow or just UV excited filaments. This
might be the reason why HH\,777 was not mentioned by Ogura et al.
(\cite{2002AJ....123.2597O}) although it is clearly visible in their image.

\begin{figure}[t]
\centering
\resizebox{\hsize}{!}{\includegraphics[angle=-90,width=6.5cm]{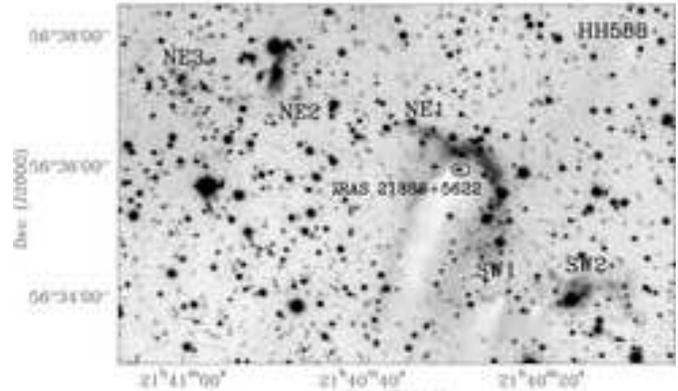}}
\caption{[SII] image of the HH\,588 region. The newly detected emission knots
HH\,588\,NE3 are seen in the upper left corner.}
\label{hh588}
\end{figure}

We also detect the HH\,354 and HH\,588 flows from IRAS\,22051+5848 and
IRAS\,21388+5622, respectively. There is an additional emission feature
1$\arcmin\!\!$.4 east of \object{HH\,588\,NE2}, consisting of two fuzzy blobs
(for positions see Table\,\ref{hhcoord}). This object was slightly outside the
field shown in Ogura et al. (\cite{2002AJ....123.2597O}). The knot-like
appearance of this object suggests an HH object (HH\,588\,NE3) connected to the
HH\,588 flow, rather than being a UV-excited emission feature (see
Fig.\,\ref{hh588}). None of the optical emission features of HH\,588 is
detected in our H$_2$ images.

\begin{figure}[t]
\centering
\resizebox{\hsize}{!}{\includegraphics[angle=-90,width=6.5cm]{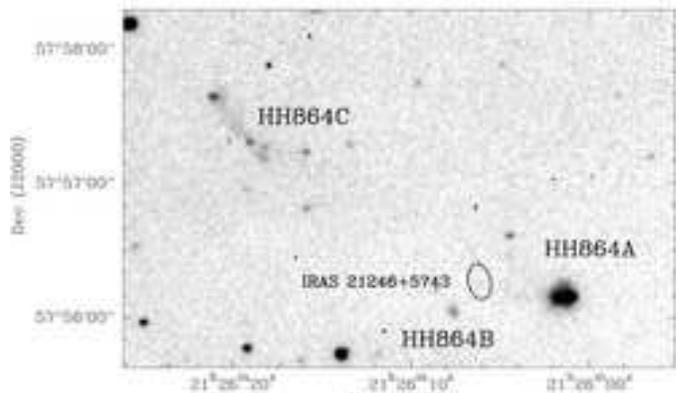}}
\caption{Our [SII] image of the outflow from IC\,1396\,W (HH\,864) shows clearly
some of the brightest emission knots, previously detected in H$_2$ by Froebrich
\& Scholz (\cite{2003A&A...407..207F}).}
\label{hh864}
\end{figure}

We detected two new groups of HH objects in our optical field. We found the
H$_2$ flow from IC\,1396\,W also in [SII] and call it HH\,864 (see
Fig.\,\ref{hh864}). The bright H$_2$ knot 2-j (Froebrich \& Scholz
\cite{2003A&A...407..207F}) has a [SII] counterpart (HH\,864\,A) which shows a
bow shape with two maxima in the emission. The two knots (2-b and 2-d) in the
north-eastern lobe are visible in the optical emission line also (HH\,864\,C).
Knot 5 (50\arcsec\, east of 2-l) has an optical counterpart as well
(HH\,864\,B). It is still not clear if this emission knot is connected to the
IRAS\,21246+5743 source or another (unknown) source in the IC\,1396\,W globule.

\begin{figure}[t]
\centering
\resizebox{\hsize}{!}{\includegraphics[angle=-90,width=6.5cm]{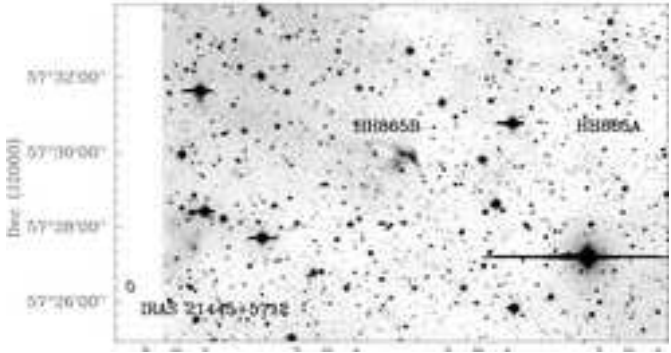}}
\vspace{0.5cm}
\caption{The new flow HH\,865 in the [SII] filter, heading away from the
possible driving source IRAS\,21445+5712.}
\label{hh865}
\end{figure}

We find in our optical field a giant flow emerging from the
IRAS\,21445+5712 source, which is also called IC\,1396\,E. There are two bright
bow shocks heading to the north-west of the source, HH\,865\,A and
HH\,865\,B (see Fig.\,\ref{hh865}). The distance of about 0.2\degr\, of the
terminating bow HH\,865\,A from the source makes this flow 2.6\,pc in length
in one lobe (assuming a distance of 725\,pc). No  counterflow is found in
our images. The globule was also observed in H$_2$. The closer of the two bow
shocks is unfortunately just outside our NIR field and hence could not be
detected. There is no H$_2$ emission in our image that might be connected to
this flow. Only a few very faint features are found south and east of the IRAS
source. These are UV excited emission coinciding with cloud borders. 

In IRAS\,21312+5736 we see some faint H$_2$ filaments that might be UV excited.
The field of IRAS\,21324+5716 is full of bright H$_2$ emission features
coinciding with the optical emission. This applies also for the neighbouring
fields IRAS\,21346+5714 and IRAS\,21352+5715. No H$_2$ features are detected
near IRAS\,21354+5823, IRAS\,21428+5802, IRAS\,21539+5821 and
IRAS\,22051+5848. 

\section{Summary and discussion}
\label{discuss}
 
NIR observations of IC\,1396 in conjunction with extinction maps obtained from
2MASS data reveal star forming activity and a large number of globules in this
region. Twenty five globules were identified using our extinction maps and the
list of Schwartz et al. (\cite{1991ApJ...370..263S}). Four of them were
previously uncatalogued in the SIMBAD database. In all but four cases the
masses (or at least lower limits) of the globules could be determined. Also the
size could be measured properly for all but seven objects. 

For nine globules observed deeper than 2MASS in this work, the content of
heavily reddened objects was derived by means of colour-colour diagrams.
Five of these globules exhibit a rich population of red objects. At least
half of these objects are good candidates for young stellar objects, the
remaining half is probably contaminated by reddened background stars. The
five globules with many red objects include the targets with the highest
extinction values, suggesting a correlation of the strength of the star
formation activity with the mass of the globule.

Star formation in small globules is often thought to be strongly influenced by
the radiation pressure of a nearby bright star. It was therefore investigated
how the globule properties in IC1396 depend on the distance from the O star 
HD\,206267. The masses of the globules show a clear positive correlation with
the  distance from this star. We conclude that evaporation due to
photo-ionisation affects the mass distribution of the globules around
HD\,206267. Our data are consistent with a scenario in which the radiation
pressure from the O star regulates the star forming activity (expressed as
density of young sources) in the globules, in the sense that the radiation
pressure compresses the gas and thus leads to enhanced star formation.

Our optical data lead to the discovery of several new HH objects. These are the
counterpart of the previously known H$_2$ emission of the flow from IC\,1396\,W
and a new parsec scale flow from IC\,1396\,E. Further a new emission knot
belonging to the known HH\,588 object was discovered.

\begin{acknowledgements}

We thank J.\,Woitas for providing observing time during his TLS run for this
project. 
D.\,Froebrich and G.C.\,Murphy received funding by the Cosmo Grid project,
funded by the Program for Research in Third Level Institutions administered by
the Irish Higher Education Authority under the National Development Plan and
with assistance from the European Regional Development Fund.
A.\,Scholz work was partially funded by Deutsche Forschungsgemeinschaft (DFG)
grants Ei409/11-1 and 11-2 to J.\,Eisl\"offel.
This publication makes use of data products from the Two Micron All Sky Survey,
which is a joint project of the University of Massachusetts and the Infrared
Processing and Analysis Center/California Institute of Technology, funded by
the National Aeronautics and Space Administration and the National Science
Foundation.
This research has made use of the SIMBAD database, operated at CDS, Strasbourg,
France.

\end{acknowledgements}

\begin{appendix}

\section{Reliability of instrumental photometry}

\label{reliability}

Several tests were executed to evaluate the reliability of the instrumental
photometry. Since we observed each globule at least twice (and six of them 
three times), we can compare the instrumental magnitudes from different 
mosaics of the same region to look for systematic mismatches. With one 
exception (see below), we found good agreement within $\pm 0.05$\,mag. The 
photometry from mosaics which were taken under non-photometric conditions  (see
Sect. \ref{nirobs}) shows the highest deviations, as expected. These 
deviations, however, affect all broadband mosaics equally, thus they are not 
visible in the colours. For one field (IRAS\,21539+5821) we see large 
systematic photometry mismatches between all mosaics, which are also visible in
the colours. The most probable explanation is variable weather conditions,
e.g. cirrus clouds which we did not recognise during the observations. We
excluded  this field from all further analysis.

As mentioned in Sect. \ref{nirobs}, all broad-band images were obtained at low
airmass; the maximum airmass difference is 0.45. Since the NIR extinction 
coefficient on Calar Alto is usually below 0.1\,mag/airmass (Hopp \&
Fern\'andez \cite{hf02}), the differential extinction offsets between the
instrumental  magnitudes of different fields are safely below 0.05\,mag. More
important, the  extinction coefficient in the NIR is nearly wavelength
independent. Therefore, differential extinction does not significantly affect
the colours of our targets. Thus, we did not perform an extinction correction.
Another test of our photometry was enabled by field overlap of the mosaics
around IRAS\,21346+5714 and IRAS\,21352+5715. Again we found no significant
offsets between the colours of the targets which were detected in both fields.
For these reasons, we conclude that the instrumental colours of the targets
observed under photometric conditions are reliable. 

If available, the deep, stacked mosaics of the globules were calibrated by 
measuring the magnitude offsets to a mosaic obtained under photometric 
conditions and applying these offsets to the instrumental magnitudes of the 
deep mosaics. Thus, the photometry of the deep mosaics is now directly
comparable with all other data obtained under photometric conditions, in
particular with the standard star main sequence.

\section{Colour-colour diagrams and absolute calibration}

\label{cc_cal}

For the colour-colour diagrams, we used the photometry from the deep mosaics if
more than one mosaic is available, otherwise the photometry from the single
mosaic. We only consider objects with errors below 0.2\,mag. There are, 
however, only very few objects (typically below 1\%) with larger errors.

The bulk of the IC\,1396 objects in the diagrams coincides with the late-type
end of the main sequence. There is one exception -- Fig.\,\ref{cc5} -- which
will be discussed below. This is expected because most stars are late-type.
Moreover, giant stars also concentrate towards the late-type end of the main
sequence. Since our main sequence ends at spectral type K8, it  is not
surprising that there are many field objects above the upper end of this 
sequence. These could be M-type dwarf or giant stars. The bulk of the field 
objects is (again with one exception) always in the same position in each
diagram,  which makes us confident that the colours from the different fields
can be compared  with each other. 

There is one field (IRAS\,22051+5848, see Fig. \ref{cc5}), where the bulk of
the  datapoints is clearly offset by about 0.2\,mag in (H-K) and 0.2\,mag in
(J-H), i.e. roughly in the direction of the reddening vector. This field shows
large-scale structures of nebulosity and large voids without any objects,
indicative of strong extinction. Thus, in this field background objects are
probably significantly reddened because their light must pass through extended
regions of dust. The 0.2\,mag offset in both colours then corresponds to a mean
visual extinction of A$_{\rm V} \approx 2$\,mag. 

We are now interested in detecting objects whose position in the colour-colour
diagram indicates significant intrinsic reddening, i.e whose position is 
clearly shifted in the direction of the extinction path or even below these
lines. The following criteria were used to select such reddened objects:

a) If an object is detected in all three filters, its H-K colour should exceed
0.5\,mag: H-K\,$>$\,0.5\,mag. (For the IRAS\,22051+5848 field, we require
H-K\,$>$\,0.7\,mag,  because all objects in this field show a 0.2\,mag shift in
H-K, see above.) Objects that satisfy this condition are shown as filled
squares in the diagrams.

b) If the object is only detected in H and K, we demand again that
H-K\,$>$\,0.5\,mag (or $>$\,0.7\,mag for IRAS\,22051+5848). These objects are
marked with an 'arrow up', and their J-H colour is a lower limit estimated
from the H-band photometry and our sensitivity limit in the J-band.

c) If the object is only detected in K, we can only determine a lower limit for
the H-K colour by subtracting the K-band photometry from the sensitivity  limit
in the H-band. We require that this lower limit is $>$\,0.5\,mag (or
$>$\,0.7\,mag for IRAS\,22051+5848). These objects are shown with an 'arrow to
the right', and their J-H colours are chosen arbitrarily.

We selected all objects that satisfy one of these conditions and examined
these targets in the images. We rejected all objects that are clearly not
star-shaped (i.e. galaxies or H$_2$ emission knots), as well as spurious
detections (i.e. cosmics or spikes of bright stars). The remaining objects are
good candidates for objects with intrinsic reddening in the globules.  This 
selection might be incomplete, since some young objects could have
H-K\,$<$\,0.5, and thus would fall out of our selection criteria. As noted
above, the colour-colour diagrams contain only objects with errors below
0.2\,mag. With larger error bars a reliable separation between reddened and
unreddened  objects is not possible. In most cases, however, the objects with
errors  $>$\,0.2\,mag do not appear strongly reddened.

After this selection process, the globules clearly fall in two groups: Four of 
them show only very few reddened objects, i.e. their number is $\le$\,10. The 
remaining five globules harbour more than 20 reddened objects, in two cases 
(IRAS\,21346+5714, IRAS\,22051+5848) the number of reddened objects is
larger than 50. For these five globules we show the colour-colour diagrams in 
Figs. \ref{cc1}-\ref{cc5}. 

As noted above, the colours in the diagrams are instrumental. We determined,
however, the offsets to the photometric system of the 2MASS catalogue, which is
available online\footnote{see {\it http://www.ipac.caltech.edu/2mass}}. For all
detected objects, we searched for counterparts in the 2MASS database. In each
field, several hundred common objects were identified. For these objects, we
determined the average offset $\Delta$ between 2MASS photometry and our
instrumental magnitudes in J, H, and K. These offsets (called $\Delta_J$,
$\Delta_H$, $\Delta_K$ in the following) include a zero-point as well as
extinction correction. 

The offsets in a certain band are not constant for all globules; they differ by
as much as 0.3\,mag, depending on the weather conditions during the
observations (see above). The difference between the offsets, in particular the
values of  $\Delta_J - \Delta_H$ and $\Delta_H - \Delta_K$, are comparable for
all globules, confirming again that the relative colours are reliable. We
obtain  $\Delta_J - \Delta_H = 0.03 \pm 0.06$ and $\Delta_H - \Delta_K = 0.29
\pm 0.08$. Thus, our colour-colour diagrams can be roughly transformed into the
2MASS photometric system by adding 0.3\,mag to the H-K values. We note,
however, that this transformation does not include colour terms, which means
that the offsets are probably not constant for all colours, in particular not
for the main sequence and the reddened targets (see Froebrich \& Scholz
\cite{2003A&A...407..207F}). This was the main reason why we kept the diagrams
in the instrumental colours, as described above.

The absolute calibration also allows us to estimate the sensitivity limit of
our  survey, i.e. the faintest objects for which a 5$\sigma$ detection is
possible. We reach about 17.0\,mag in J, 16.5\,mag in H, and 16.0\,mag in K.
For comparison, the 2MASS catalogue for this region contains objects down to
16.9\,mag in J, 16.1\,mag in H, and 15.6\,mag in K. Our images are thus 
significantly deeper than the 2MASS catalogue for this region, particularly  in
the H- and K-band. 

\section{Extinction map determination}

\label{exmade}

A determination of absolute magnitudes of the stars within the globules, as
well as an estimation of the dust mass in the globules, requires a measurement
of the extinction. As discussed above, the assumption of a uniform Galactic
extinction is not valid in our field. There are local extinction enhancements
connected to the globules and even within these small clouds the dust and so
the extinction is not uniformly distributed. Hence we need to determine the
local extinction enhancements for our globules, compared to the neighbouring
star field. These extinction enhancements will be used to estimate the mass of
the globules by adopting a distance of 750\,pc. Measuring the extinction from
colour-colour diagrams is relatively difficult due to the photometric errors
and the large scatter of the main sequence (see e.g. our
Figs.\,\ref{cc1}-\ref{cc5}). Hence this method only allows a rough estimate of
at best $\pm$\,3\,mag of the optical extinction averaged over the whole globule
and is in particular not able to determine the extinction/mass distribution
within the globules. A more accurate method is to create accumulative star
counts (Wolf diagrams) as described e.g. in Kiss et al.
(\cite{2000A&A...363..755K}). Since we expect high extinction values within the
globules (A$_{\rm V}$\,$\ge$\,5\,mag), creating these diagrams from NIR
observations obviously is the best choise. The 2MASS database provides an ideal
basis for this purpose. 

\begin{figure}[t]
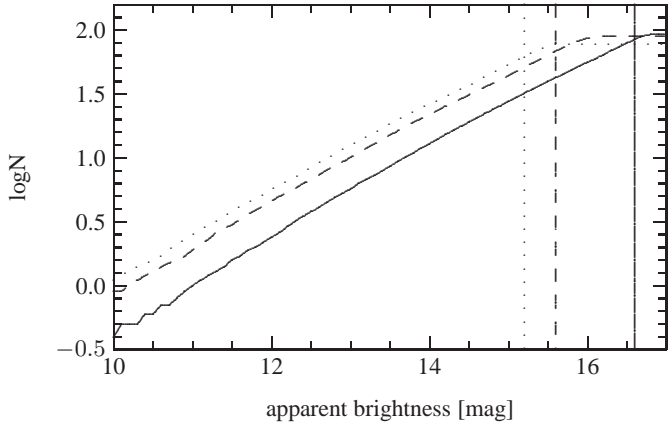

\beginpicture
\setcoordinatesystem units <10.5mm,17mm> point at 0 0
\setplotarea x from 10 to 17 , y from -0.5 to 2.2
\setsolid
\plot
10.0  -0.39794   
10.1  -0.30103   
10.2  -0.30103   
10.3  -0.30103   
10.4  -0.22184   
10.5  -0.22184   
10.6  -0.15490   
10.7  -0.15490   
10.8  -0.09691   
10.9  -0.04575   
11.0   0.00000   
11.1   0.04139   
11.2   0.07918   
11.3   0.11394   
11.4   0.14612   
11.5   0.20412   
11.6   0.23044   
11.7   0.27875   
11.8   0.30103   
11.9   0.34242   
12.0   0.38021   
12.1   0.41497   
12.2   0.46239   
12.3   0.50515   
12.4   0.54406   
12.5   0.57978   
12.6   0.61278   
12.7   0.65321   
12.8   0.69019   
12.9   0.72427   
13.0   0.75587   
13.1   0.79934   
13.2   0.83250   
13.3   0.86332   
13.4   0.89762   
13.5   0.93449   
13.6   0.97312   
13.7   1.00860   
13.8   1.04139   
13.9   1.07918   
14.0   1.11059   
14.1   1.14613   
14.2   1.18184   
14.3   1.21484   
14.4   1.24797   
14.5   1.28103   
14.6   1.31387   
14.7   1.34635   
14.8   1.37840   
14.9   1.40993   
15.0   1.44248   
15.1   1.47422   
15.2   1.50651   
15.3   1.53782   
15.4   1.56820   
15.5   1.59879   
15.6   1.62839   
15.7   1.65801   
15.8   1.68842   
15.9   1.71850   
16.0   1.74819   
16.1   1.77597   
16.2   1.80754   
16.3   1.84011   
16.4   1.86982   
16.5   1.90037   
16.6   1.92840   
16.7   1.95376   
16.8   1.96661   
16.9   1.97313   
17.0   1.97313   
/
\setdashes
\plot  
10.0  -0.04575
10.1  -0.04575
10.2   0.00000
10.3   0.04139
10.4   0.07918
10.5   0.11394
10.6   0.14612
10.7   0.17609
10.8   0.20412
10.9   0.23044
11.0   0.27875
11.1   0.32221
11.2   0.36172
11.3   0.41497
11.4   0.44715
11.5   0.47712
11.6   0.51851
11.7   0.55630
11.8   0.59106
11.9   0.62324
12.0   0.66275
12.1   0.69897
12.2   0.73239
12.3   0.76342
12.4   0.79934
12.5   0.83250
12.6   0.86923
12.7   0.90309
12.8   0.93951
12.9   0.96848
13.0   1.00432
13.1   1.04139
13.2   1.07555
13.3   1.11059
13.4   1.14301
13.5   1.17898
13.6   1.21219
13.7   1.24551
13.8   1.27875
13.9   1.31175
14.0   1.34242
14.1   1.37291
14.2   1.40483
14.3   1.43297
14.4   1.46240
14.5   1.49415
14.6   1.52375
14.7   1.55509
14.8   1.58433
14.9   1.61490
15.0   1.64640
15.1   1.67761
15.2   1.70927
15.3   1.74036
15.4   1.77232
15.5   1.80209
15.6   1.83569
15.7   1.86629
15.8   1.89542
15.9   1.91960
16.0   1.93702
16.1   1.94645
16.2   1.94939
16.3   1.95036
16.4   1.95085
16.5   1.95085
16.6   1.95085
16.7   1.95085
16.8   1.95085
16.9   1.95085
17.0   1.95085
/
\setdots
\plot  
10.0   0.04139
10.1   0.07918
10.2   0.11394
10.3   0.14612
10.4   0.17609
10.5   0.20412
10.6   0.23044
10.7   0.27875
10.8   0.32221
10.9   0.36172
11.0   0.39794
11.1   0.43136
11.2   0.47712
11.3   0.50515
11.4   0.54406
11.5   0.57978
11.6   0.62324
11.7   0.65321
11.8   0.68124
11.9   0.71600
12.0   0.75587
12.1   0.79239
12.2   0.81954
12.3   0.85125
12.4   0.88649
12.5   0.92427
12.6   0.95904
12.7   0.99563
12.8   1.02938
12.9   1.06446
13.0   1.09691
13.1   1.13033
13.2   1.16732
13.3   1.19866
13.4   1.23553
13.5   1.26717
13.6   1.29885
13.7   1.33041
13.8   1.35984
13.9   1.39094
14.0   1.41996
14.1   1.44716
14.2   1.47712
14.3   1.50651
14.4   1.53908
14.5   1.56937
14.6   1.59988
14.7   1.63347
14.8   1.66370
14.9   1.69373
15.0   1.72673
15.1   1.75891
15.2   1.79169
15.3   1.82413
15.4   1.85309
15.5   1.87506
15.6   1.88705
15.7   1.89098
15.8   1.89098
15.9   1.89098
16.0   1.89098
16.1   1.89098
16.2   1.89098
16.3   1.89098
16.4   1.89098
16.5   1.89098
16.6   1.89098
16.7   1.89098
16.8   1.89098
16.9   1.89098
17.0   1.89098
/

\setsolid
\plot 16.6 2.2 16.6 -0.5 /
\setdashes
\plot 15.6 2.2 15.6 -0.5 /
\setdots
\plot 15.2 2.2 15.2 -0.5 /
\setsolid

\axis left label {\begin{sideways}logN\end{sideways}}
ticks in long numbered from -0.5 to 2.0 by 0.5
      short unlabeled from -0.5 to 2.2 by 0.1 /
\axis right label {}
ticks in long unlabeled from -0.5 to 2.0 by 0.5
      short unlabeled from -0.5 to 2.2 by 0.1 /
\axis bottom label {apparent brightness [mag]}
ticks in long numbered from 10 to 17 by 2
      short unlabeled from 10 to 17 by 0.5 /
\axis top label {}
ticks in long unlabeled from 10 to 17 by 2
      short unlabeled from 10 to 17 by 0.5 /
\endpicture
\caption{\label{wolfdiag} Example of a Wolf diagram for the three filters
(J - solid; H - dashed; K - dotted) at $\alpha$\,=\,21$^{\rm h}$52$^{\rm
m}$00$^{\rm s}$ $\delta$\,=\,57\degr00\arcmin00\arcsec\, (J2000). The
accumulated number of stars N is averaged over a 1\degr\,x\,1\degr field and
normalised to an area of nine square arcminutes. The vertical lines correspond
to the completeness limit in JHK (line type as for the Wolf diagrams) at which
we measured the extinction. The mean slope $<$X$>$ varies with the filter (0.36
for J and 0.34 for H and K) and also with the position on the sky. As can be
seen the slope is also not constant, due to extinction from the ISM (the slope
in J ranges from 0.38 (at 11 to 12\,mag) to 0.30 (at 15 to 16\,mag)).}
\end{figure}

Using star counts to estimate the extinction is based on several assumptions:
1) The stars are equally distributed and all apparent voids or less densely
populated regions are caused by extinction. 2) All stars possess the same
absolute brightness. 3) The completeness limit of the input catalogue does not
depend on the position in the sky. Since in general these assumptions are not
valid, we performed our star counts in a way that ensures as small as possible
errors due to the assumptions made. 1) As reference or control field we choose
a 1$^\circ$\,x\,1$^\circ$ field around the position where the stars are
counted. This running average ensures that large scale variations in the
average star numbers due to Galactic position or large scale structures are
corrected. 2) The box size in which we count the stars for each position was
varied between 1\arcmin$\times$1\arcmin\, and 5\arcmin$\times$5\arcmin. In
boxes larger than 3\arcmin$\times$3\arcmin\, of the order of 100\,stars are
enclosed on average. This ensures that we enclose a 'typical' sample of stars
in our box, that represents the 'average' star very well. 3) The completeness
limit of the 2MASS catalogue varies over the field. Star counts where hence
only performed down to a magnitude to which the catalogue is complete over the
whole field (J=16.6, H=15.6, K=15.2).

\begin{figure}[t]
\centering
\includegraphics[angle=-90,width=9.cm]{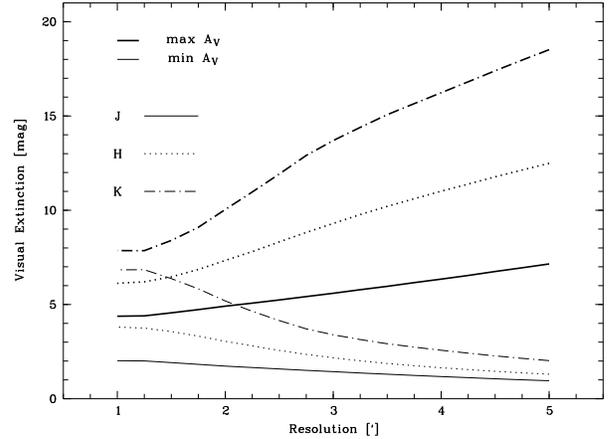}
\caption{\label{ranges} Minimum and maximum values of optical extinction that
could be traced with our extinction maps obtained from the 2MASS data in JHK,
using different box sizes. A$_{\rm V}$ values are determined from
extinction in JHK using the conversion factors given in Mathis et al.
(\cite{1990eism.conf...63M}).}
\end{figure}

In order to investigate not only the extinction distribution in our globules
but also the large scale dust distribution in the entire Cep\,OB\,2 region, we
created an extiction map of the whole field, ranging from 21$^{\rm h}\!\!$.3 to
22$^{\rm h}\!\!$.2 in R.A. and from 56$^\circ$ to 60$^\circ$ in DEC (J2000). In
Fig.\,\ref{obsfield} we show an example of the extinction map obtained from the
J-band 2MASS data.  Since we had a large amount of data to process we decided
to parallelise the problem. We wrote an MPI-parallel C++ code in order to get
maximum performance from our 32-node hyperthreaded cluster of the so-called
"Beowulf" type. It took 13 hours to process all the data on this cluster -
without the parallelisation of the problem it would have taken of the order of
60 times as long (about 33 days).

For the star counts we selected all 2MASS objects in our field with a
signal-to-noise ratio of larger than five (quality flag A, B, or C). With this
selection criteria we extracted 841908, 783474, and 635478 objects in JHK,
respectively.  This converts to an average density of stars of about
1\,star per 20\arcsec\,x\,20\arcsec\, field. Hence, we performed the star
counts every 20\arcsec\, in order to obtain the final extinction map. This
method ensures a maximum gain of information from the 2MASS catalogue.
Figure\,\ref{wolfdiag} shows an example of a Wolf diagram in each of the three
filters at $\alpha$\,=\,21$^{\rm h}$52$^{\rm m}$00$^{\rm s}$
$\delta$\,=\,57\degr00\arcmin00\arcsec\, (J2000). In order to find the best
compromise between box-size and extinction regime that can be traced, the
box-sizes were varied in steps of 0\arcmin$\!\!$.25. Depending on the box-size
the average number of stars in the boxes varies and hence we can trace
different extinction regimes. In Fig.\,\ref{ranges} we plot the minimum and
maximum value of traceable extinction (converted to A$_{\rm V}$) with our
method in IC\,1396, depending on the chosen box-size and the three 2MASS
filters. The lower limit (A$_\lambda^{\rm min}$) is calculated from the three
sigma noise level of the determined extinction in the whole Cep\,OB\,2 field.
This limit can be transformed into a ratio F$_\sigma$/F$_{\rm back}$ by
\begin{equation} 
{\rm A}_\lambda^{\rm min} = - \frac{1}{\rm X} * \log{ \left( 1
- \frac{{\rm F}_\sigma}{{\rm F}_{\rm back}} \right)}. 
\end{equation} 
F$_\sigma$ is the noise of the number of stars in the star count map. If
this number of stars is missing in the presence of F$_{\rm back}$ stars in the
comparison field, an apparent extinction of A$_\lambda^{\rm min}$ is detected.
The factor X is the slope in the Wolf-diagram. Note that this factor, as well
as the number of background stars varies with the position and filter. The
maximum tracable extinction A$_\lambda^{\rm max}$ can now be determined by
assuming that only F$_\sigma$ stars are present.
\begin{equation} 
{\rm
A}_\lambda^{\rm max} = - \frac{1}{\rm X} * \log{ \left( \frac{{\rm
F}_\sigma}{{\rm F}_{\rm back}} \right)} 
\end{equation} 
This is a conservative assumption, because F$_\sigma$ represents actually
the noise of the number of background and foreground stars, but only the latter
determines the maximum of tracable extinction. Note: If we assume that the
scatter of foreground stars accounts only for half of F$_\sigma$, the maximum
tracable extinction values increase by about 4, 6, and 10\,mag optical
extinction for JHK, respectively (assuming X\,=\,0.3). We selected a box-size
of 3\arcmin\,$\times$\,3\arcmin\, to perform all our measurements. The lower
limit (A$_\lambda^{\rm min}$) is also comparable to the error of our determined
extinction values. As can be seen in Fig.\,\ref{ranges} this is much lower than
the estimated accuracy of $\pm$\,3\,mag when the extinction is inferred from
the colour-colour diagrams. 

\end{appendix}


\begin{thebibliography}{}

\bibitem[1998]{1998A&A...337..403B}
Baraffe, I., Chabrier, G. Allard, F. \& Hauschildt, P.H. 1998, A\&A, 337, 403

\bibitem[2002]{2002ApJ...573..246B}
Beltr\'an, M.T., Girart, J.M., Estalella, R., Ho, P.T.P. \& Palau, A. 2002, ApJ, 573, 246

\bibitem[1996]{1996A&AS..117..393B}
Bertin, E. \& Arnouts, S. 1996, A\&AS, 117, 393

\bibitem[2000]{2000ApJ...542..464C} 
Chabrier, G., Baraffe, I., Allard, F. \& Hauschildt, P. 2000, ApJ, 542, 464

\bibitem[2001]{2001A&A...376..271C}
Codella, C., Bachiller, R., Nisini, B., Saraceno, P. \& Testi, L. 2001, A\&A, 376, 271

\bibitem[2002]{2002ApJ...577..798D}
De Vries, C.H., Narayanan, G. \& Snell, R.L. 2002, ApJ, 577, 798

\bibitem[1990]{1990A&A...233..190D}
Duvert, G., Cernicharo, J., Bachiller, R. \& G\'omez-Gonz\'alez, J. 1990, A\&A, 233, 190

\bibitem[2003]{2003A&A...407..207F}
Froebrich, D. \& Scholz, A. 2003, A\&A, 407, 207

\bibitem[2000]{2000AJ....120.1085G}
Gizis, J.E., Monet, D.G., Reid, I.N., Kirkpatrick, J.D., Liebert, J. \& Williams, R.J. 2000, AJ, 120, 1085

\bibitem[1993]{1993SPIE.1946..605H}
Herbst, T.M., Beckwith, S.V., Birk, C., Hippler, S., McCaughrean, M.J., Mannucci, F. \& Wolf, J. 1993, SPIE, 1946, 605

\bibitem[1995]{1995A&A...299..464H}
Hessman, F.V., Beckwith, S.V.W., Bender, R., Eisl\"offel, J., G\"otz, W. \& Guenther, E. 1995, A\&A, 299, 464

\bibitem[2002]{hf02}
Hopp, U. \& Fern\'andez, M. 2002, Calar Alto Newsletter Nr. 4

\bibitem[2000]{2000A&A...363..755K}
Kiss, C,. T\'oth, L.V., Mo\'or, A., Sato, F., Nikolic, S. \& Wouterloot, J.G.A. 2000, A\&A, 363, 755

\bibitem[1990]{1990eism.conf...63M}
Mathis, J.S. 1990, ARA\&A, 28, 37

\bibitem[1979]{1979A&A....75..345M}
Matthews, H.I. 1979, A\&A, 75, 345

\bibitem[1989]{1989PASJ...41.1073N}
Nakano, M., Tomita, Y., Ohtani, H., Ogura, K. \& Sofue, Y. 1989, PASJ, 41, 1073

\bibitem[2001]{2001A&A...376..553N}
Nisini, B., Massi, F., Vitali, F., Giannini, T., et al. 2001, A\&A, 376, 553

\bibitem[2002]{2002AJ....123.2597O}
Ogura, K., Sugitani, K. \& Pickles, A. 2002, AJ, 123, 2597

\bibitem[1995]{1995ApJ...447..721P}
Patel, N.A., Goldsmith, P.F., Snell, R.L., Hezel, T. \& Xie, T. 1995, ApJ, 447, 721

\bibitem[2004]{2004ApJS..154..385R}
Reach, W.T., Rho, J., Young, E., et al. 2004, ApJS, 154, 385

\bibitem[1983]{1983A&A...117..183R}
Reipurth, B. 1983, A\&A, 117, 183

\bibitem[2003]{2003ApJ...593L..47R}
Reipurth, B., Armond, T., Raga, A. \& Bally, J. 2003, ApJ, 593, 47

\bibitem[2001]{2001ARA&A..39..403R}
Reipurth, B. \& Bally, J. 2001, ARA\&A, 39, 403

\bibitem[2003]{2003A&A...409..523R} 
Robin, A. C., Reyl\'e, C., Derriere, S. \& Picaud, S. 2003, A\&A, 409, 523

\bibitem[1991]{1991ApJ...370..263S}
Schwartz, R.D., Wilking, B.A. \& Giulbudagian, A.L. 1991, ApJ, 370, 263

\bibitem[1993]{1993ApJ...404..247S}
Serabyn, E., Guesten, R. \& Mundy, L. 1993, ApJ, 404, 247
        
\bibitem[1995]{1995ApJ...450..512S}
Stanford, S.A., Eisenhardt, P.R.M. \& Dickinson, M. 1995, ApJ, 450, 512

\bibitem[1991]{1991ApJS...77...59S}
Sugitani, K., Fukui, Y. \& Ogura, K. 1991, ApJS, 77, 59

\bibitem[1997]{1997ApJ...486L.141S}
Sugitani, K., Morita, K.-I., Nakano, M., Tamura, M. \& Ogura, K. 1997, ApJ, 486, 141

\bibitem[1996]{1996A&A...309..581W}
Weikard, H., Wouterloot, J.G.A., Castets, A., Winnewisser, G. \& Sugitani, K. 1996, A\&A, 309, 581

\bibitem[1993]{1993AJ....106..250W}
Wilking, B., Mundy, L., McMullin, J., Hezel, T. \& Keene, J. 1993, AJ, 106, 205

\bibitem[1984]{1984ApJ...286..718W}
Walborn, N.R. \& Panek, R.J. 1984, ApJ, 286, 718

\end{thebibliography}
\end{document}